\documentclass[11pt]{article}
\usepackage[dvipdfmx]{graphicx}
\usepackage{bmpsize}
\usepackage{graphicx}
\usepackage{amsmath}
\usepackage{amssymb}
\usepackage{latexsym}
\usepackage{overpic}
\usepackage{here}
\usepackage{pdflscape}
\usepackage{bm}
\usepackage{empheq}
\usepackage{arydshln}
\usepackage{ulem}
\newcount\num \num=0
\def\NUM{\global\advance\num by 1
\the\num}
\newcommand{\ds}{\displaystyle}
\newtheorem{theorem}{\bf Theorem}
\newtheorem{lemma}{\bf Lemma}
\newtheorem{example}{\bf Example}

\newtheorem{remark}{\bf Remark}
\title{\bf\Large On a Proof of the Convergence Speed of a Second-order Recurrence Formula in the Arimoto-Blahut Algorithm}
\author{Kenji Nakagawa
\thanks{Nagaoka University of Technology, E-mail: {\tt nakagawa@nagaokaut.ac.jp}}\\
\and Yoshinori Takei
\thanks {Nippon Sport Science University
}\\
\and Shin-ichiro Hara
\thanks{Nagaoka University of Technology}
}
\date{}
\begin{document}
\maketitle

\vspace{-2mm}

\begin{abstract}
In \cite{nak5} (Nakagawa,\,et.al.,\,IEEE Trans.\,IT, 2021), we investigated the convergence speed of the Arimoto-Blahut algorithm. In \cite{nak5}, the convergence of the order $O(1/N)$ was analyzed by focusing on the second-order nonlinear recurrence formula consisting of the first- and second-order terms of the Taylor expansion of the defining function of the Arimoto-Blahut algorithm. However, in \cite{nak5}, an infinite number of inequalities were assumed as a ``conjecture,'' and proofs were given based on the conjecture. In this paper, we report a proof of the convergence of the order $O(1/N)$ for a class of channel matrices without assuming the conjecture. The correctness of the proof will be confirmed by several numerical examples.

\end{abstract}
\section{Introduction}
In \cite{nak5}, we showed that in the Arimoto-Blahut algorithm, the speed at which the input probability distribution $\bm\lambda^N$ converges to the capacity achieving distribution $\bm\lambda^\ast$ includes exponential convergence and convergence of the order $O(1/N)$. In \cite{nak5}, the convergence of the order $O(1/N)$ was analyzed by focusing on the second-order nonlinear recurrence formula consisting of the first- and second-order terms of the Taylor expansion of the defining function of the Arimoto-Blahut algorithm. However, in \cite{nak5}, an infinite number of inequalities were assumed as a ``conjecture'' and proofs were given based on the conjecture. In this paper, we report a proof of convergence of the order $O(1/N)$ for a class of channel matrices without assuming the conjecture. A key idea of the proof is to examine the continuous graph obtained by connecting the discrete graph of a sequence of numbers defined by the second-order recurrence formula with a line segment. The correctness of the proof will be confirmed by several numerical examples.

However, this paper still assumes some finite number of inequalities. It is a future challenge to reduce these assumptions as much as possible.

\section{Arimoto-Blahut algorithm}
\subsection{Channel matrix and channel capacity}
Consider a discrete memoryless channel $X\rightarrow Y$ with the input source $X$ and the output source $Y$. Let ${\cal X}=\{x_1,\dots,x_m\}$ be the input alphabet and ${\cal Y}=\{y_1,\dots,y_n\}$ be the output alphabet. The conditional probability that the output symbol $y_j$ is received when the input symbol $x_i$ is transmitted is denoted by $P^i_j=P(Y=y_j|X=x_i),\,i=1,\dots,m, j=1,\dots,n,$ and the row vector $P^i$ is defined by $P^i=(P^i_1,\dots,P^i_n),\,i=1,\dots,m$. The channel matrix $\Phi$ is defined by
\begin{align}
\label{eqn:thechannelmatrix}
\Phi=\begin{pmatrix}
\,P^1\,\\
\vdots\\
\,P^m\,
\end{pmatrix}
=\begin{pmatrix}
\,P^1_1 & \dots & P^1_n\,\\
\vdots & & \vdots\\
\,P^m_1 & \dots & P^m_n\,
\end{pmatrix}.
\end{align}
We assume that for any $j\,(j=1,\dots,n)$, there exists at least one $i\,(i=1,\dots,m)$ with $P^i_j>0$. This assumption means that there are no useless output symbols.

The set of input probability distributions on the input alphabet ${\cal X}$ is denoted by $\Delta({\cal X})\equiv\{\bm\lambda=(\lambda_1,\dots,\lambda_m)|\lambda_i\geq0,i=1,\dots,m,\sum_{i=1}^m\lambda_i=1\}$. The interior of $\Delta({\cal X})$ is denoted by $\Delta({\cal X})^\circ\equiv\{\bm\lambda=(\lambda_1,\dots,\lambda_m)\in\Delta({\cal X})\,|\,\lambda_i>0,\,i=1,\dots,m\}$. Similarly, the set of output probability distributions on the output alphabet ${\cal Y}$ is denoted by $\Delta({\cal Y})\equiv\{Q=(Q_1,\dots,Q_n)|Q_j\geq0,j=1,\dots,n,\sum_{j=1}^nQ_j=1\}$, and its interior $\Delta({\cal Y})^\circ$ is similarly defined as $\Delta({\cal X})^\circ$.

Let $Q=\bm\lambda\Phi$ be the output distribution for an input distribution $\bm\lambda\in\Delta(\cal X)$ and write its components as $Q_j=\sum_{i=1}^m\lambda_iP^i_j,\,j=1,\dots,n$. Then, the mutual information is defined by $I(\bm\lambda,\Phi)=\sum_{i=1}^m\sum_{j=1}^n\lambda_iP^i_j\log\left({P^i_j}/{Q_j}\right)$, where $\log$ is the natural logarithm. The channel capacity $C$ is defined by
\begin{align}
\label{eqn:Cdefinition}
C=\max_{\bm\lambda\in\Delta({\cal X})}I(\bm\lambda,\Phi).
\end{align}
The unit of $C$ is nat/symbol. 

The Kullback-Leibler divergence $D(Q\|Q')$ for two probability distributions $Q=(Q_1,\dots,Q_n)$, $Q'=(Q'_1,\dots,Q'_n)$ is defined \cite{csi1} by
\begin{align}
D(Q\|Q')=\sum_{j=1}^nQ_j\log\ds\frac{Q_j}{Q'_j}.
\end{align}

An important proposition for investigating the convergence speed of the Arimoto-Blahut algorithm is the Kuhn-Tucker condition on the input distribution $\bm\lambda=\bm\lambda^\ast$ that achieves the maximum of (\ref{eqn:Cdefinition}).

\medskip

\noindent {\bf Theorem A\ \it \text{\rm(}Kuhn-Tucker condition \text{\rm\cite{cov}}\text{\rm)} In the maximization problem \text{\rm(\ref{eqn:Cdefinition})}, a necessary and sufficient condition for the input distribution $\bm\lambda^\ast=(\lambda^\ast_1,\dots,\lambda^\ast_m)\in\Delta({\cal X})$ to achieve the maximum is that there is a certain constant $\tilde{C}$ with
\begin{align}
\label{eqn:Kuhn-Tucker}
D(P^i\|\bm\lambda^\ast\Phi)\left\{\begin{array}{ll}=\tilde{C}, & {\mbox{\rm for}}\ i\ {\mbox{\rm with}}\ \lambda^\ast_i>0,\\
\leq \tilde{C}, & {\mbox{\rm for}}\ i\ {\mbox{\rm with}}\ \lambda^\ast_i=0.
\end{array}\right.
\end{align}
In \text{\rm(\ref{eqn:Kuhn-Tucker})}, $\tilde{C}$ is equal to the channel capacity $C$.\rm

\medskip

Since the Kuhn-Tucker condition is a necessary and sufficient condition, all the information about the capacity-achieving input distribution $\bm\lambda^\ast$ can be derived from this condition.
\subsection{Definition of the Arimoto-Blahut algorithm}
A sequence $\{\bm\lambda^N=(\lambda^N_1,\dots,\lambda^N_m)\}_{N=0,1,\dots}\subset\Delta({\cal X})$ of input distributions is defined by the Arimoto-Blahut algorithm as follows \cite{ari}, \cite{bla}. First, let $\bm\lambda^0=(\lambda^0_1,\dots,\lambda^0_m)$ be an initial distribution taken in $\Delta(\cal X)^\circ$, i.e., $\lambda^0_i>0,\,i=1,\dots,m$. Then, the Arimoto-Blahut algorithm is given by the recurrence formula
\begin{align}
\lambda^{N+1}_i=\ds\frac{\lambda^N_i\exp D(P^i\|\bm\lambda^N\Phi)}{\ds\sum_{i'=1}^m\lambda^N_{i'}\exp D(P^{i'}\|\bm\lambda^N\Phi)},\ N=0,1,\dots,\ i=1,\dots,m.\label{eqn:arimotoalgorithm}
\end{align}
\subsection{Function from $\Delta({\cal X})$ to $\Delta({\cal X})$}
Let $F_i(\bm\lambda)$ be the defining function of the Arimoto-Blahut algorithm (\ref{eqn:arimotoalgorithm}), i.e., 
\begin{align}
F_i(\bm\lambda)=\ds\frac{\lambda_i\exp D(P^i\|\bm\lambda\Phi)}{\ds\sum_{i'=1}^m\lambda_{i'}\exp D(P^{i'}\|\bm\lambda\Phi)},\,i=1,\dots,m.\label{eqn:Arimotofunction}
\end{align}
Define $F(\bm\lambda)\equiv(F_1(\bm\lambda),\dots,F_m(\bm\lambda))$. Then, $F(\bm\lambda)$ is a differentiable function from $\Delta(\cal X)$ to $\Delta(\cal X)$, and (\ref{eqn:arimotoalgorithm}) is represented by $\bm\lambda^{N+1}=F(\bm\lambda^N)$.

In this paper, for the analysis of the convergence speed, we assume 
\begin{align}
{\rm rank}\,\Phi=m.\label{eqn:rankmdefinition}
\end{align}
\begin{lemma}\text{\rm \cite{nak5}}
\label{lem:capacity_achieving_lambda_is_the_fixed_point}
The capacity-achieving input distribution $\bm\lambda^\ast$ is the fixed point of the function $F(\bm\lambda)$. That is, $F(\bm\lambda^\ast)=\bm\lambda^\ast$.
\end{lemma}

\medskip

The sequence $\left\{\bm\lambda^N\right\}_{N=0,1,\dots}$ of the Arimoto-Blahut algorithm converges to the fixed point $\bm\lambda^\ast$, i.e., $\bm\lambda^N\to\bm\lambda^\ast,\,N\to\infty$.
We will investigate the convergence speed by using the Taylor expansion of $F(\bm\lambda)$ about $\bm\lambda=\bm\lambda^\ast$.
\subsection{Convergence speed of $\bm\lambda^N\to\bm\lambda^\ast$}
Now, we define two kinds of convergence speeds for investigating $\bm\lambda^N\to\bm\lambda^\ast$. 
\begin{itemize}
\item[$\bullet$] Exponential convergence\\
$\bm\lambda^N\to\bm\lambda^\ast$ is the {\it exponential convergence} if
\begin{align}
\|\bm\lambda^N-\bm\lambda^\ast\|<K\cdot(\theta)^N,\ K>0,\ 0\leq\theta<1,\ N=0,1,\dots,
\end{align}
where $\|\bm\lambda\|$ denotes the Euclidean norm, i.e., $\|\bm\lambda\|=\left(\lambda_1^2+\dots+\lambda_m^2\right)^{1/2}$, and $(\theta)^N$ denotes $\theta$ to the power $N$.
\item[$\bullet$] Convergence of the order $O(1/N)$ \\
$\bm\lambda^N\to\bm\lambda^\ast$ is the {\it convergence of the order $O(1/N)$} if 
\begin{align}
\lim_{N\to\infty}N\left(\lambda^N_i-\lambda^\ast_i\right)=K_i\neq0,\,i=1,\dots,m.
\end{align}
\end{itemize}
\subsection{Type of index}
Now, we classify the indices $i\,(i=1,\dots,m)$ in the Kuhn-Tucker condition (\ref{eqn:Kuhn-Tucker}) in more detail into the following 3 types.
\begin{align}
\label{eqn:Kuhn-Tucker2}
D(P^i\|\bm\lambda^\ast\Phi)\left\{\begin{array}{ll}=C, & {\mbox{\rm for}}\ i\ {\mbox{\rm with}}\ \lambda^\ast_i>0\ \mbox{\rm [type-I]},\\
=C, & {\mbox{\rm for}}\ i\ {\mbox{\rm with}}\ \lambda^\ast_i=0\ \mbox{\rm [type-II]},\\
<C, & {\mbox{\rm for}}\ i\ {\mbox{\rm with}}\ \lambda^\ast_i=0\ \mbox{\rm [type-III]}.
\end{array}\right.
\end{align}
Let us define the sets of indices as follows:
\begin{align}
&{\rm all\ the\ indices:}\ {\cal I}\equiv\{1,\dots,m\},\label{eqn:allset}\\
&\mbox{\rm type-I\ indices:}\ {\cal I}_{\rm I}\equiv\{1,\dots,m_1\},\label{eqn:type1set}\\
&\mbox{\rm type-II\ indices:}\ {\cal I}_{\rm II}\equiv\{m_1+1,\dots,m_1+m_2\},\label{eqn:type2set}\\
&\mbox{\rm type-III\ indices:}\ {\cal I}_{\rm III}\equiv\{m_1+m_2+1,\dots,m\}.\label{eqn:type3set}
\end{align}
We have $|{\cal I}|=m$, $|{\cal I}_{\rm I}|=m_1$, $|{\cal I}_{\rm II}|=m_2$, $|{\cal I}_{\rm III}|=m-m_1-m_2\equiv m_3$, ${\cal I}={\cal I}_{\rm I}\cup{\cal I}_{\rm II}\cup{\cal I}_{\rm III}$ and $m=m_1+m_2+m_3$. For any channel matrix, ${\cal I}_{\rm I}$ is not empty and $|{\cal I}_{\rm I}|=m_1\geq2$, but ${\cal I}_{\rm II}$ and ${\cal I}_{\rm III}$ may be empty for some channel matrices.
\subsection{Related Works}
The previous studies \cite{ari}, \cite{ari2}, \cite{yu} investigate the exponential convergence. In \cite{ari}, \cite{ari2}, \cite{yu}, it is proved that if the capacity achieving input distribution $\bm\lambda^\ast=(\lambda^\ast_1,\dots,\lambda^\ast_m)$ satisfies $\lambda^\ast_i>0,\,\,i=1\,\dots,m$, then the convergence of $\bm\lambda^N\to\bm\lambda^\ast$ is exponential. In this case, all indices are type-I and there are no type-II indices nor type-III indices. When there are no type-II indices, the speed of convergence is determined by the Jacobian matrix of the defining function $F(\bm\lambda)$ of the Arimoto-Blahut algorithm, and the analysis is not very difficult.

On the other hand, in \cite{nak5}, we showed that when there are type-II indices, the convergence speed of the Arimoto-Blahut algorithm is not determined by the Jacobian matrix alone, and the Hessian matrix must also be examined. Then, the convergence of the order $O(1/N)$ was investigated using the second-order nonlinear recurrence formula obtained by the Jacobian and Hessian matrices.

\section{Taylor expansion of $F(\bm\lambda)$ about $\bm\lambda=\bm\lambda^\ast$}
The convergence speed of the Arimoto-Blahut algorithm will be investigated by the Taylor expansion of $F(\bm\lambda)$ about the fixed point $\bm\lambda=\bm\lambda^\ast$.

The Taylor expansion of the function $F(\bm\lambda)=(F_1(\bm\lambda),\dots,F_m(\bm\lambda))$ about $\bm\lambda=\bm\lambda^\ast$ is
\begin{align}
F(\bm\lambda)=F(\bm\lambda^\ast)+(\bm\lambda-\bm\lambda^\ast)J(\bm\lambda^\ast)+\ds\frac{1}{2!}(\bm\lambda-\bm\lambda^\ast)H(\bm\lambda^\ast)\,^t(\bm\lambda-\bm\lambda^\ast)+o\left(\|\bm\lambda-\bm\lambda^\ast\|^2\right),\label{eqn:Taylortenkai1}
\end{align}
where ${^t}\bm\lambda$ is the transpose of $\bm\lambda$. In (\ref{eqn:Taylortenkai1}), $J(\bm\lambda^\ast)$ is the Jacobian matrix of $F(\bm\lambda)$ in $\bm\lambda=\bm\lambda^\ast$, i.e.,
\begin{align}
J(\bm\lambda^\ast)=\left(\left.\dfrac{\partial F_i}{\partial\lambda_{i'}}\right|_{\bm\lambda=\bm\lambda^\ast}\right)_{i',i=1,\dots,m}.
\end{align}
Further, in (\ref{eqn:Taylortenkai1}), $H(\bm\lambda^\ast)\equiv(H_1(\bm\lambda^\ast),\dots,H_m(\bm\lambda^\ast))$, where $H_i(\bm\lambda^\ast)$ is the Hessian matrix of $F_i(\bm\lambda)$ at $\bm\lambda=\bm\lambda^\ast$, i.e.,
\begin{align}
H_i(\bm\lambda^\ast)=\left(\left.\dfrac{\partial^2F_i}{\partial\lambda_{i'}\partial\lambda_{i''}}\right|_{\bm\lambda=\bm\lambda^\ast}\right)_{i',i''=1,\dots,m},
\end{align}
and $(\bm\lambda-\bm\lambda^\ast)H(\bm\lambda^\ast)\,^t(\bm\lambda-\bm\lambda^\ast)$ is an abbreviated expression of the $m$-dimensional row vector
\begin{align}
\left((\bm\lambda-\bm\lambda^\ast)H_1(\bm\lambda^\ast)\,^t(\bm\lambda-\bm\lambda^\ast),\dots,(\bm\lambda-\bm\lambda^\ast)H_m(\bm\lambda^\ast)\,^t(\bm\lambda-\bm\lambda^\ast)\right).
\end{align}

In (\ref{eqn:Taylortenkai1}), put $\bm\lambda=\bm\lambda^N$, $\bm\mu^N\equiv\bm\lambda^N-\bm\lambda^\ast$, then from $F(\bm\lambda^\ast)=\bm\lambda^\ast$ and $F(\bm\lambda^N)=\bm\lambda^{N+1}$, we have
\begin{align}
\bm\mu^{N+1}=\bm\mu^NJ(\bm\lambda^\ast)+\ds\frac{1}{2!}\bm\mu^NH(\bm\lambda^\ast)\,{^t}\bm\mu^N+o\left(\|\bm\mu^N\|^2\right).\label{eqn:Taylortenkai3}
\end{align}

From $\bm\lambda^N\to\bm\lambda^\ast$ we have $\bm\mu^N\to\bm0$. The convergence of $\bm\mu^N\to\bm0,\,N\to\infty$ is analyzed based on the Taylor expansion (\ref{eqn:Taylortenkai3}). We write the components of $\bm\mu^N$ as $\bm\mu^N=(\mu^N_1,\dots,\mu^N_m)$, then $\mu^N_i=\lambda^N_i-\lambda^\ast_i,\,i=1,\dots,m$.

\subsection{Eigenvalues of the Jacobian matrix $J(\bm\lambda^\ast)$}
The Jacobian matrix has the following form \cite{nak5}:
\begin{align}
&\hspace{0mm}J(\bm\lambda^\ast)\equiv\begin{pmatrix}
\,J^{\rm I} & O & O\,\\[1mm]
\,\ast & J^{\rm II} & O\,\\[1mm]
\,\ast & O & J^{\rm III}
\end{pmatrix},\label{eqn:J1AJ2}\\
&J^{\rm I}\equiv\left(\partial F_i/\partial\lambda_{i'}|_{\bm\lambda=\bm\lambda^\ast}\right)_{i,i'\in{\cal I}_{\rm I}}\in{\mathbb R}^{m_1\times m_1},\nonumber\\
&J^{\rm II}\equiv\left(\partial F_i/\partial\lambda_{i'}|_{\bm\lambda=\bm\lambda^\ast}\right)_{i,i'\in{\cal I}_{\rm II}}\in{\mathbb R}^{m_2\times m_2},\nonumber\\
&J^{\rm III}\equiv\left(\partial F_i/\partial\lambda_{i'}|_{\bm\lambda=\bm\lambda^\ast}\right)_{i,i'\in{\cal I}_{\rm III}}\in{\mathbb R}^{m_3\times m_3},\nonumber\\
&O\ \mbox{\rm denotes\ the\ all-zero\ matrix\ of\ appropriate\ size,}\nonumber\\
&*\ \mbox{\rm denotes an appropriate matrix}.\nonumber
\end{align}

Let $J(\bm\lambda^\ast)$ have eigenvalues $\{\theta_1,\dots,\theta_m\}\equiv\{\theta_i\,|\,i\in{\cal I}\}$, then from (\ref{eqn:J1AJ2}), the eigenvalues of $J^{\rm I}$, $J^{\rm II}$, $J^{\rm III}$ are $\{\theta_i\,|\,i\in{\cal I}_{\rm I}\}$, $\{\theta_i\,|\,i\in{\cal I}_{\rm II}\}$ $\{\theta_i\,|\,i\in{\cal I}_{\rm III}\}$, respectively. For each eigenvalue, the following evaluations are obtained.
\begin{theorem}{\rm \cite{nak5}}
\label{the:eigenvaluesofJ1J2J3}
The eigenvalues of $J^{\rm I}$ satisfy $0\leq\theta_i<1,\,i\in{\cal I}_{\rm I}$. The eigenvalues of $J^{\rm II}$ satisfy $\theta_i=1,\,i\in{\cal I}_{\rm II}$, more precisely, $J^{\rm II}=I$ \text{\rm(}the identity matrix\text{\rm)} $\in\mathbb{R}^{m_2\times m_2}$. The eigenvalues of $J^{\rm III}$ satisfy $0<\theta_i<1,\,i\in{\cal I}_{\rm III}$.
\end{theorem}
\section{Analysis of the convergence speed of the order $O(1/N)$}
From Theorem \ref{the:eigenvaluesofJ1J2J3}, when there is no type-II index, i.e., ${\cal I}_{\rm II}=\emptyset$, all the eigenvalues of $J(\bm\lambda^\ast)$ satisfy $0\leq\theta_i<1$, hence the speed of convergence of the recurrence formula $\bm\lambda^{N+1}=F(\bm\lambda^N)$ is determined only by $J(\bm\lambda^\ast)$, and it is an exponential convergence \cite{ari}, \cite{ari2}, \cite{nak5}, \cite{yu}.

In contrast, when ${\cal I}_{\rm II}\neq\emptyset$, the maximum eigenvalue of $J(\bm\lambda^\ast)$ is $1$, so the convergence speed is not determined by $J(\bm\lambda^\ast)$ alone, and the Hessian matrix $H_i(\bm\lambda^\ast)$ must be analyzed.

Therefore, in the following, we assume
\begin{align}
{\cal I}_{\rm II}\neq\emptyset,
\end{align}
and consider the convergence speed of the order $O(1/N)$.

Now, we consider the types-I and -III indices together and reorder the indices so that ${\cal I}_{\rm I}\cup{\cal I}_{\rm III}=\{1,\dots,m'\}$ and ${\cal I}_{\rm II}=\{m'+1,\dots,m\}$. We have $m_2=m-m'$ and $|{\cal I}_{\rm II}|=m_2$. We define $\bm\mu^N_{\rm I,III}\equiv(\mu^N_1,\dots,\mu^N_{m'})$ by $\bm\mu^N_{\rm II}\equiv(\mu^N_{m'+1},\dots,\mu^N_m)$.

In the Jacobian matrix (\ref{eqn:J1AJ2}), by changing the order of $J^{\rm II}$ and $J^{\rm III}$, we have
\begin{align}
J(\bm\lambda^\ast)=\begin{pmatrix}
\,J^{\rm I} & O & O\,\\[1mm]
\,\ast & J^{\rm III} & O\,\\[1mm]
\,\ast & O & J^{\rm II}
\end{pmatrix}.
\end{align}
Then, by defining 
\begin{align}
\label{eqn:J'JIJIII}
J'\equiv\begin{pmatrix}
\,J^{\rm I} & O \,\\[1mm]
\,\ast & J^{\rm III}
\end{pmatrix},
\end{align}
we have
\begin{align}
\label{eqn:JJ'JII}
J(\bm\lambda^\ast)=\begin{pmatrix}
\,J' & O \,\\[1mm]
\,\ast & J^{\rm II}
\end{pmatrix}.
\end{align}
By Theorem \ref{the:eigenvaluesofJ1J2J3}, the eigenvalues of $J(\bm\lambda^\ast)$ are $\theta_i,\,i=1,\dots,m$, and then the eigenvalues of $J'$ are $\theta_i,\,i=1,\dots,m'$ with $0\leq\theta_i<1$, and $J^{\rm II}=I\in{\mathbb R}^{m_2\times m_2}$.

\subsection{Second-order recurrence formula of $\bm\mu^N_{\rm II}$}
Now, let $\bm a_i$ be a right eigenvector of $J(\bm\lambda^\ast)$ for $\theta_i$ and define
\begin{align}
A\equiv(\bm a_1,\dots,\bm a_m)\in{\mathbb R}^{m\times m}.\label{eqn:a1...am}
\end{align}
$J(\bm\lambda^\ast)$ is diagonalizable under the assumption that $\theta_i\neq\theta_{i'}$ for $i\in\cal{I}_{\rm I}$ and $i'\in\cal{I}_{\rm III}$ (see Lemma 5 in \cite{nak5}). Then, by choosing the eigenvectors $\bm a_1,\dots,\bm a_m$ appropriately, we can make $A$ a regular matrix (see \cite{sat},\,p.161,\,Example 4).

For $i=m'+1,\dots,m$, define
\begin{align}
\bm{e}_i=(0,\dots,0,\stackrel{i\,\text{th}}{\stackrel{\vee}{1}}\hspace{-1mm},\ 0,\dots,0)\in{\mathbb R}^m.
\end{align}
Then, because $J^{\rm II}=I$, we can take
\begin{align}
{\bm a}_i={^t}\bm{e}_i,\,i=m'+1,\dots,m.\label{eqn:eigenvectorfor1}
\end{align}
Therefore, we have
\begin{align}
A&=\left(\bm a_1,\dots,\bm a_{m'},{^t}\bm e_{m'+1},\dots,{^t}\bm e_m\right)\\
&=\begin{pmatrix}
\begin{array}{l:c}
\begin{matrix}\hspace{1mm}a_{11}&\hspace{1mm}\dots&\hspace{3mm}a_{m'1}\\
\hspace{3.5mm}\vdots&&\vdots\\
\hspace{2mm}a_{1m'}&\hspace{2mm}\dots&\hspace{2mm}a_{m'm'}
\end{matrix}&O\\[7mm]
\hdashline\\[-4mm]
\begin{matrix}a_{1,m'+1}&\dots&a_{m',m'+1}\\\vdots&&\vdots\\a_{1m}&\dots&a_{m'm}\end{matrix}&
\begin{matrix}1&\dots&0\\\vdots&\ddots&\vdots\\0&\dots&1\end{matrix}
\end{array}
\end{pmatrix}\\
&\equiv\begin{pmatrix}A_1&O\\A_2&I\end{pmatrix},\label{eqn:A1OA2I}
\end{align}
where
\begin{align}
A_1&\equiv\begin{pmatrix}a_{11}&\dots&a_{m'1}\\\vdots&&\vdots\\a_{1m'}&\dots&a_{m'm'}\end{pmatrix}\in{\mathbb R}^{m'\times m'},\\
A_2&\equiv\begin{pmatrix}a_{1,m'+1}&\dots&a_{m',m'+1}\\\vdots&&\vdots\\a_{1m}&\dots&a_{m'm}\end{pmatrix}\in{\mathbb R}^{m_2\times m'}.\label{eqn:A1A2definition}
\end{align}
Because $A$ is regular, $A_1$ is also regular by (\ref{eqn:A1OA2I}). Then, under the assumptions in \cite{nak5}, we have
\begin{align}
\bm\mu^N_{\rm I,III}=-\bm\mu^N_{\rm II}A_2A_1^{-1}.\label{eqn:bmmuN{I,III}A1+bmmuN{II}A2=bm0}
\end{align}
See Eq. (99) in \cite{nak5}.
\begin{remark}
$A,A_1,A_2$ depend on the choice of eigenvectors, but $A_2A_1^{-1}$ does not depend on their choice.\end{remark}

Consider the following recurrence formula consisting of the first- and second-order terms of the Taylor expansion (\ref{eqn:Taylortenkai3}):
\begin{align}
\bm\mu^{N+1}=\bm\mu^NJ(\bm\lambda^\ast)+\ds\frac{1}{2!}\bm\mu^NH(\bm\lambda^\ast)\,{^t}\bm\mu^N.\label{eqn:linearcombination}
\end{align}

From (\ref{eqn:bmmuN{I,III}A1+bmmuN{II}A2=bm0}), (\ref{eqn:linearcombination}) and the Hessian matrix which is calculated in \cite{nak5}, the following recurrence formula for $\bm\mu^N_{\rm II}$ is obtained.%
\begin{theorem}\text{\rm (Theorem 7 in \cite{nak5})}
\label{the:1/Norderkihonzenkashiki}
$\bm\mu^N_{\rm II}=(\mu_{m'+1}^N,\dots,\mu_m^N)$ satisfies a second-order recurrence formula of the following form:
\begin{align}
\mu^{N+1}_i&=\mu^N_i+\mu^N_i\ds\sum_{i'=m'+1}^mr_{ii'}\mu^N_{i'},\ i=m'+1,\dots,m,\label{eqn:1/Norderzenkashiki}
\end{align}
where $r_{ii'}$ are determined by $J(\bm\lambda^\ast),\,A_1,\,A_2$, and $H(\bm\lambda^\ast)$.
\end{theorem}

The purpose of the study is to show that the sequence $\{\mu_i^N\}_{N=0,1,\dots}$ defined by (\ref{eqn:1/Norderzenkashiki}) converges to $0$ at the speed of the order $O(1/N)$ and to calculate $\lim_{N\to\infty}N\mu_ i^N$. To do so, we will reformulate (\ref{eqn:1/Norderzenkashiki}) in a canonical form and we make some assumptions.
\subsection{Canonical form of (\ref{eqn:1/Norderzenkashiki}) and some assumptions}
\label{sec:canonicalform}
About (\ref{eqn:1/Norderzenkashiki}), define $R\equiv(r_{ii'})_{i,i'=m'+1,\dots,m}$, $\bm1\equiv(1,\dots,1)\in{\mathbb R}^{m_2},\,(m_2=m-m')$. Now, we assume the following assumptions (i)-(iv) on $R$:
\begin{align*}
&\text{\rm(i)}\ r_{ii'}\leq0,\,i,i'=m'+1,\dots,m,\\
&\text{\rm(ii)}\ R\ \text{\rm is\ a\ regular\ matrix},\\
&\text{\rm(iii)}\ \text{\rm for}\ \bm\sigma=-\bm1(^tR)^{-1}\,{\rm with}\ \bm\sigma=(\sigma_{m'+1},\dots,\sigma_m),\ \text{\rm we\ have}\ \sigma_i>0,\,i=m'+1,\dots,m,\\
&\text{\rm(iv)}\ -r_{ii}\sigma_i>1/2,\,i=m'+1,\dots,m.
\end{align*}

\bigskip

Let us define $\nu_i^N\equiv\mu_i^N/\sigma_i,\,i=m'+1,\dots,m,\ q_{ii'}\equiv-r_{ii'}\sigma_{i'},\,i,i'=m'+1,\dots,m$, then
\begin{align}
&\nu_i^{N+1}=\nu_i^N-\nu_i^N\sum_{i'=1}^mq_{ii'}\nu_{i'}^N,\,i=m'+1,\dots,m,\label{eqn:canonical_recursion_formula}\\
&q_{ii'}\geq0,\,i,i'=m'+1,\dots,m,\label{eqn:nonnegative}\\
&\sum_{i'=m'+1}^mq_{ii'}=1,\,i=m'+1,\dots,m,\label{eqn:sumisone}\\
&q_{ii}>1/2,\,i=m'+1,\dots,m.\label{eqn:diagonalyuui}
\end{align}

From (\ref{eqn:nonnegative}) and (\ref{eqn:sumisone}), ${\bm q}_i\equiv(q_{i,m'+1},\dots,q_{im})$ is a probability vector. We call (\ref{eqn:canonical_recursion_formula})-(\ref{eqn:diagonalyuui}) a canonical form of (\ref{eqn:1/Norderzenkashiki}) under the assumptions (i)-(iv).

In \cite{nak5}, in addition to the above assumptions (i)-(iv), we assumed, through a lot of numerical examples, the inequalities
\begin{align}
\nu_{m'+1}^N\geq\dots\geq\nu_m^N,\ \exists N_0\geq0,\ \forall N\geq N_0,\label{eqn:daishouhozon}
\end{align}
by reordering the indices in ${\cal I}_{\rm II}$ if necessary. In \cite{nak5}, we called (\ref{eqn:daishouhozon}) a conjecture. However, these are an infinite number of inequalities and cannot be checked numerically. In this paper, we will prove the convergence speed of $\bm\mu^N\to\bm0$ without assuming (\ref{eqn:daishouhozon}), but assuming only (i)-(iv).

\section{Second-order simultaneous recurrence formulas}
\label{sec:shuusokushoumei}
(\ref{eqn:canonical_recursion_formula})-(\ref{eqn:diagonalyuui}) are complicated, so for simplicity, we will change the variable names and indices in (\ref{eqn:canonical_recursion_formula})-(\ref{eqn:diagonalyuui}) as follows, and add an initial condition.

Consider the following second-order simultaneous recurrence formulas for $h$ variables $z_1^N,\dots,z_h^N$. For $i=1,\dots,h$, 
\begin{align}
&z_i^{N+1}=z_i^N-z_i^N\sum_{i'=1}^hq_{ii'}z_{i'}^N,\label{eqn:recurrecne_formula}\\
&\text{\rm initial\ condition:}\ 0<z_i^0\leq1/2,\label{eqn:0<xii0leq1/2}\\
&{\bm q}_i\equiv(q_{i1},\dots,q_{ih})\ \text{\rm is\ a\ probability\ vector},\label{eqn:prob_v}\\
&\text{\rm diagonal\ dominance\ condition:}\ q_{ii}>1/2.\label{eqn:diagonal}
\end{align}

Now, let us define
\begin{align}
w^N_i\equiv\sum_{i'=1}^hq_{ii'}z_{i'}^N,\ N=0,1,\dots,\ i=1,\dots,h.\label{eqn:defoftau}
\end{align}
Then, by (\ref{eqn:defoftau}), we can write (\ref{eqn:recurrecne_formula}) as
\begin{align}
z_i^{N+1}=z_i^N-z_i^Nw^N_i,\ i=1,\dots,h.
\end{align}

For each $N\geq0$, let us define $\overline{i}(N)$ as the index that achieves $\max\left(z_1^N,\dots,z_h^N\right)$ and put $\overline{Z}^N\equiv z_{\,\overline{i}(N)}^N$, i.e.,
\begin{align}
\overline{Z}^N=z_{\,\overline{i}(N)}^N=\max\left(z_1^N,\dots,z_h^N\right).\label{eqn:max_of_z^N}
\end{align}
Similarly, for $N\geq0$, define $\underline{i}(N)$ as the index that achieves $\min\left(z_1^N,\dots,z_h^N\right)$ and put $\underline{Z}^N\equiv z_{\,\underline{i}(N)}^N$, i.e.,
\begin{align}
\underline{Z}^N=z_{\,\underline{i}(N)}^N=\min\left(z_1^N,\dots,z_h^N\right).\label{eqn:min_of_z^N}
\end{align}
\subsection{Convergence speed of the Order $O(1/N)$}
The purpose of the study is to prove the following Theorem \ref{the:main} for $\{z_i^N\}$ that satisfies (\ref{eqn:recurrecne_formula})-(\ref{eqn:diagonal}).
\begin{theorem}
\label{the:main}
\ $\ds\lim_{N\to\infty}Nz_i^N=1,\ i=1,\dots,h$.
\end{theorem}
To that end, we prove the following Theorem \ref{the:hasamiuchi}.
\begin{theorem}
\label{the:hasamiuchi}
\ $\ds\lim_{N\to\infty}N\overline{Z}^N=1,\ \ds\lim_{N\to\infty}N\underline{Z}^N=1$.
\end{theorem}

Theorem \ref{the:main} follows from Theorem \ref{the:hasamiuchi} and the squeeze theorem.

\medskip

Now, let us prove Theorem \ref{the:hasamiuchi}.

\begin{lemma}
\label{lem:0iNleq12}
\ $0<z_i^N\leq1/2,\ N=0,1,\dots,\ i=1,\dots,h$.
\end{lemma}
\noindent{\bf Proof:} We prove it by mathematical induction. For $N=0$, the assertion holds by (\ref{eqn:0<xii0leq1/2}). Assuming that the assertion holds for $N$ and noting $1/2\leq1-\sum_{i'=1}^hq_{ii'}z_{i'}^N<1$ by (\ref{eqn:prob_v}), we have $0<z_i^{N+1}\leq1/2$.\hfill$\blacksquare$

\bigskip

\begin{lemma}
\label{lem:0<w_i^N<=1/2}
\ $0<w_i^N\leq1/2,\ N=0,1,\dots,\ i=1,\dots,h$.
\end{lemma}
\noindent{\bf Proof:} The assertion holds by (\ref{eqn:defoftau}), (\ref{eqn:prob_v}) and Lemma \ref{lem:0iNleq12}.\hfill$\blacksquare$

\bigskip

\begin{lemma}
\label{lem:xiNmonotone}
\ For $i=1,\dots,h$, the sequence $\{z_i^N\},\,N=0,1,\dots$ is strictly decreasing.
\end{lemma}
\noindent{\bf Proof:}\ Because $z_i^N-z_i^{N+1}=z_i^N\sum_{i'=1}^hq_{ii'}z_{i'}^N>0$ holds by Lemma \ref{lem:0iNleq12}.\hfill$\blacksquare$

\bigskip

\begin{lemma}
\label{lem:limNtoinftyxiN=0}
$\ds\lim_{N\to\infty}z_i^N=0,\ i=1,\dots,h$.
\end{lemma}
\noindent{\bf Proof:}\ $z_i^\infty\equiv\lim_{N\to\infty}z_i^N\geq0$ exists by Lemma \ref{lem:0iNleq12} and Lemma \ref{lem:xiNmonotone}. Then, $z_i^\infty=z_i^\infty-z_i^\infty\sum_{i'=1}^hq_{ii'}z_{i'}^\infty$ holds by (\ref{eqn:recurrecne_formula}), hence we have $z_i^\infty=0$ by (\ref{eqn:diagonal}).\hfill$\blacksquare$

\bigskip

We will use the following theorem in our proof.

\noindent{\bf Theorem B}\ (\cite{ahl},\,\it p.\text{\rm 37},\,Exercise \text{\rm 2)}\ If $\ds\lim_{N\to\infty}a_N=a$, then $\ds\lim_{N\to\infty}\ds\frac{1}{N}\left(a_0+a_1+\dots+a_{N-1}\right)=a$.\rm

\begin{lemma}
\label{lem:liminfNtoinftyNoverlineZNgeq1}
\ $\ds\liminf_{N\to\infty}N\overline{Z}^N\geq1$.
\end{lemma}
\noindent{\bf Proof:}\ We have
\begin{align}
\overline{Z}^{N+1}&=z_{\,\overline{i}(N+1)}^{N+1}\\
&\geq z_{\,\overline{i}(N)}^{N+1}\\
&=z_{\,\overline{i}(N)}^N-z_{\,\overline{i}(N)}^N\ds\sum_{i'=1}^hq_{\,\overline{i}(N)i'}z_{i'}^N\\
&\geq z_{\,\overline{i}(N)}^N-z_{\,\overline{i}(N)}^N\ds\sum_{i'=1}^hq_{\,\overline{i}(N)i'}z_{\,\overline{i}(N)}^N\\
&=\overline{Z}^N-\left(\overline{Z}^N\right)^2\ ({\rm by}\ (\ref{eqn:prob_v})).\label{eqn:overlineZN+1geq}
\end{align}
From (\ref{eqn:overlineZN+1geq}),
\begin{align}
\dfrac{1}{N}\left(\dfrac{1}{\overline{Z}^N}-\dfrac{1}{\overline{Z}^0}\right)&=\dfrac{1}{N}\sum_{l=0}^{N-1}\left(\dfrac{1}{\overline{Z}^{l+1}}-\dfrac{1}{\overline{Z}^l}\right)\\
&\leq\dfrac{1}{N}\sum_{l=0}^{N-1}\dfrac{1}{1-\overline{Z}^l}.
\end{align}
Therefore,
\begin{align}
\limsup_{N\to\infty}\dfrac{1}{N\overline{Z}^N}&\leq\limsup_{N\to\infty}\dfrac{1}{N}\sum_{l=0}^{N-1}\dfrac{1}{1-\overline{Z}^l}\\
&=\lim_{N\to\infty}\dfrac{1}{N}\sum_{l=0}^{N-1}\dfrac{1}{1-\overline{Z}^l}\label{eqn:Ahlfors1}\\
&=1,\label{eqn:limsupleq1}
\end{align}
where (\ref{eqn:Ahlfors1}) and (\ref{eqn:limsupleq1}) follow from Lemma \ref{lem:limNtoinftyxiN=0} and Theorem B.\hfill$\blacksquare$

\bigskip

\begin{lemma}
\label{lem:limsupNtoinftyNunderlineZNleq1}
\ $\ds\limsup_{N\to\infty}N\underline{Z}^N\leq1.$
\end{lemma}
\noindent{\bf Proof:}\ We have
\begin{align}
\underline{Z}^{N+1}&=z_{\,\underline{i}(N+1)}^{N+1}\\
&\leq z_{\,\underline{i}(N)}^{N+1}\\
&=z_{\,\underline{i}(N)}^N-z_{\,\underline{i}(N)}^N\ds\sum_{i'=1}^hq_{\,\underline{i}(N)i'}z_{i'}^N\\
&\leq z_{\,\underline{i}(N)}^N-z_{\,\underline{i}(N)}^N\ds\sum_{i'=1}^hq_{\,\underline{i}(N)i'}z_{\,\underline{i}(N)}^N\\
&=\underline{Z}^N-\left(\underline{Z}^N\right)^2\ ({\rm by}\ (\ref{eqn:prob_v})).\label{eqn:underlineZN+1leq}
\end{align}
From (\ref{eqn:underlineZN+1leq}),
\begin{align}
\dfrac{1}{N}\left(\dfrac{1}{\underline{Z}^N}-\dfrac{1}{\underline{Z}^0}\right)&=\dfrac{1}{N}\sum_{l=0}^{N-1}\left(\dfrac{1}{\underline{Z}^{l+1}}-\dfrac{1}{\underline{Z}^l}\right)\\
&\geq\dfrac{1}{N}\sum_{l=0}^{N-1}\dfrac{1}{1-\underline{Z}^l}.
\end{align}
Therefore,
\begin{align}
\liminf_{N\to\infty}\dfrac{1}{N\underline{Z}^N}&\geq\liminf_{N\to\infty}\dfrac{1}{N}\sum_{l=0}^{N-1}\dfrac{1}{1-\underline{Z}^l}\\
&=\lim_{N\to\infty}\dfrac{1}{N}\sum_{l=0}^{N-1}\dfrac{1}{1-\underline{Z}^l}\label{eqn:Ahlfors2}\\
&=1,\label{eqn:liminfgeq1}
\end{align}
where (\ref{eqn:Ahlfors2}) and (\ref{eqn:liminfgeq1}) follow from Lemma \ref{lem:limNtoinftyxiN=0} and Theorem B.\hfill$\blacksquare$

\bigskip

\begin{lemma}
\label{lem:xiN<dfracL1N}
For any $i=1,\dots,h$ and $N\geq1$, we have
\begin{align}
z_i^N<\dfrac{2}{N}.\label{eqn:xiN<dfracL1N}
\end{align}
\end{lemma}

\medskip

\noindent{\bf Proof:}\ We will prove it by mathematical induction. First, by the initial condition (\ref{eqn:0<xii0leq1/2}) and Lemma \ref{lem:xiNmonotone}, we have
\begin{align}
z_i^2<z_i^1<z_i^0\leq\dfrac{1}{2},
\end{align}
hence (\ref{eqn:xiN<dfracL1N}) holds for $N=1,2$.

Next, assume that (\ref{eqn:xiN<dfracL1N}) holds for $N(\geq2)$. Because the function $z-(1/2)z^2$ is monotonically increasing for $0<z\leq1$,
\begin{align}
z_i^{N+1}&=z_i^N-z_i^N\sum_{i'=1}^hq_{ii'}z_{i'}^N\\
&\leq z_i^N-q_{ii}\left(z_i^N\right)^2\\
&<z_i^N-\dfrac{1}{2}\left(z_i^N\right)^2\ \ \text{\rm(by\ (\ref{eqn:diagonal}))}\label{eqn:ziN-12ziN2}\\
&<\dfrac{2}{N}-\dfrac{1}{2}\left(\dfrac{2}{N}\right)^2\\
&<\dfrac{2}{N+1},
\end{align}
thus, (\ref{eqn:xiN<dfracL1N}) holds for $N+1$.\hfill$\blacksquare$

\subsection{Connect $z_i^N$ and $z_i^{N+1}$ with a line segment}
For $t\geq0$, we define $z_i(t)$ as the polyline obtained by connecting $z_i^0,z_i^1,\dots,z_i^N,\dots$ with line segments, i.e.,
\begin{align}
z_i(t)\equiv(N+1-t)z_i^N+(t-N)z_i^{N+1}, \ {\rm for}\ N\leq t\leq N+1,\ i=1,\dots,h.\label{eqn:polyline}
\end{align}
In particular,
\begin{align}
z_i(N)=z_i^N,\ z_i(N+1)=z_i^{N+1}.\label{eqn:zi(N)=ziNzi(N+1)=ziN+1}
\end{align}
\begin{lemma}
\label{lem:z_i(t)isstrictlydec}
$z_i(t)$ is strictly decreasing,\ $i=1,\dots,h$.
\end{lemma}
\noindent{\bf Proof:}\ The assertion holds by Lemma \ref{lem:xiNmonotone} and (\ref{eqn:polyline}).\hfill$\blacksquare$

\bigskip

Now, let us define
\begin{align}
w_i(t)\equiv\sum_{i'=1}^hq_{ii'}z_{i'}(t),\ 0\leq t<\infty,\ i=1,\dots,h.\label{eqn:definition_of_w_i(t)}
\end{align}

Next, for $t\geq0$, let us define $\overline{i}(t)$ as the index that achieves $\max\left(z_1(t),\dots,z_h(t)\right)$ and put $\overline{Z}(t)\equiv z_{\,\overline{i}(t)}(t)$, i.e.,
\begin{align}
\overline{Z}(t)=z_{\,\overline{i}(t)}(t)=\max\left(z_1(t),\dots,z_h(t)\right).\label{eqn:maz_of_x(t)}
\end{align}
Similarly, for $t\geq0$, define $\underline{i}(t)$ as the index that achieves $\min\left(z_1(t),\dots,z_h(t)\right)$ and put $\underline{Z}(t)\equiv z_{\,\underline{i}(t)}(t)$, i.e.,
\begin{align}
\underline{Z}(t)=z_{\,\underline{i}(t)}(t)=\min\left(z_1(t),\dots,z_h(t)\right).\label{eqn:min_of_x(t)}
\end{align}
In particular,
\begin{align}
\overline{Z}(N)=\overline{Z}^N,\ \underline{Z}(N)=\underline{Z}^N.\label{eqn:overlineZ(N)=overlineZ^N}
\end{align}

\bigskip

\begin{lemma}
\label{lem:overlineZt-underlineZt<L2tau-tau}
\ There exists $L_1>0$ such that for $t\geq0$,
\begin{align}
\overline{Z}(t)-\underline{Z}(t)\leq L_1\left(w_{\,\overline{i}(t)}(t)-w_{\,\underline{i}(t)}(t)\right).
\end{align}
\end{lemma}
\noindent{\bf Proof:}\ Because $t$ is fixed, we write $\overline{i}\equiv\overline{i}(t)$ and $\underline{i}\equiv\underline{i}(t)$ for the sake of simplicity. Then, we have
\begin{align}
w_{\,\overline{i}(t)}(t)-w_{\,\underline{i}(t)}(t)
&=w_{\,\overline{i}}(t)-w_{\,\underline{i}}(t)\\
&=\sum_{i'=1}^hq_{\,\overline{i}i'}z_{i'}(t)-\sum_{i'=1}^hq_{\,\underline{i}i'}z_{i'}(t)\\
&=q_{\,\overline{i}\overline{i}}z_{\,\overline{i}}(t)+\sum_{i'\neq\overline{i}}q_{\,\overline{i}i'}z_{i'}(t)-q_{\,\underline{i}\underline{i}}z_{\,\underline{i}}(t)-\sum_{i'\neq\underline{i}}q_{\,\underline{i}i'}z_{i'}(t)\\
&\geq q_{\,\overline{i}\overline{i}}z_{\,\overline{i}}(t)+\sum_{i'\neq\overline{i}}q_{\,\overline{i}i'}z_{\,\underline{i}}(t)-q_{\,\underline{i}\underline{i}}z_{\,\underline{i}}(t)-\sum_{i'\neq\underline{i}}q_{\,\underline{i}i'}z_{\,\overline{i}}(t)\\
&=q_{\,\overline{i}\overline{i}}\overline{Z}(t)+\left(1-q_{\,\overline{i}\overline{i}}\right)\underline{Z}(t)-q_{\,\underline{i}\underline{i}}\underline{Z}(t)-\left(1-q_{\,\underline{i}\underline{i}}\right)\overline{Z}(t)\\
&=\left(q_{\,\overline{i}\overline{i}}+q_{\,\underline{i}\underline{i}}-1\right)\left(\overline{Z}(t)-\underline{Z}(t)\right).\label{eqn:lem17-1}
\end{align}
Here, put 
\begin{align}
\ell\equiv\min_{i, i'}\left(q_{ii}+q_{i'i'}-1\right),
\end{align}
then $\ell>0$ by the assumption (\ref{eqn:diagonal}). Hence, the assertion of the Lemma holds by putting $L_1=\ell^{-1}$.\hfill$\blacksquare$

\bigskip

\begin{lemma}
\label{lem:Ztmonotnedecreasing}
\ $\overline{Z}(t)$ and $\underline{Z}(t)$ are strictly decreasing.
\end{lemma}

\medskip

\noindent{\bf Proof:}\ For every $i$, $z_i(t)$ is strictly decreasing by Lemma \ref{lem:z_i(t)isstrictlydec}, hence for $t_1<t_2$ we have $\overline{Z}(t_1)=z_{\,\overline{i}(t_1)}(t_1)\geq z_{\,\overline{i}(t_2)}(t_1)>z_{\,\overline{i}(t_2)}(t_2)=\overline{Z}(t_2)$. Similarly for $\underline{Z}(t)$.\hfill$\blacksquare$

\bigskip

\begin{lemma}
\label{lem:finitefinite}
\ $\ds\sum_{N=0}^\infty\left(\overline{Z}^N-\underline{Z}^N\right)<\infty$ is equivalent to $\ds\int_0^\infty\left\{\overline{Z}(t)-\underline{Z}(t)\right\}dt<\infty$.
\end{lemma}

\medskip

\noindent{\bf Proof:}\ For $N\leq t\leq N+1,\ N=0,1,\dots$, we have by Lemma \ref{lem:Ztmonotnedecreasing} and (\ref{eqn:overlineZ(N)=overlineZ^N}),
\begin{align}
&\overline{Z}^N\geq\overline{Z}(t)\geq\overline{Z}^{N+1},\label{eqn:finitefinite1}\\
&\underline{Z}^N\geq\underline{Z}(t)\geq\underline{Z}^{N+1}.\label{eqn:finitefinite2}
\end{align}
By integrating (\ref{eqn:finitefinite1}) and (\ref{eqn:finitefinite2}) on the interval $[N,N+1]$, we have
\begin{align}
&\overline{Z}^N=\int_N^{N+1}\overline{Z}^Ndt\geq\int_N^{N+1}\overline{Z}(t)dt\geq\int_N^{N+1}\overline{Z}^{N+1}dt=\overline{Z}^{N+1},\label{eqn:finitefinite3}\\
&\underline{Z}^N=\int_N^{N+1}\underline{Z}^Ndt\geq\int_N^{N+1}\underline{Z}(t)dt\geq\int_N^{N+1}\underline{Z}^{N+1}dt=\underline{Z}^{N+1}.\label{eqn:finitefinite4}
\end{align}
Then by (\ref{eqn:finitefinite3}), (\ref{eqn:finitefinite4}), we have
\begin{align}
&\overline{Z}^N\geq\int_N^{N+1}\overline{Z}(t)dt\geq\overline{Z}^{N+1},\label{eqn:finitefinite5}\\
&-\underline{Z}^{N+1}\geq-\int_N^{N+1}\underline{Z}(t)dt\geq-\underline{Z}^N\label{eqn:finitefinite6}.
\end{align}
Adding the both sides of (\ref{eqn:finitefinite5}) and (\ref{eqn:finitefinite6}), we have
\begin{align}
\overline{Z}^N-\underline{Z}^{N+1}\geq\int_N^{N+1}\left\{\overline{Z}(t)-\underline{Z}(t)\right\}dt\geq\overline{Z}^{N+1}-\underline{Z}^N.\label{eqn:finitefinite7}
\end{align}
Taking summation of the both sides of (\ref{eqn:finitefinite7}) for $N=0,1,\dots,N'$, we have, by a simple calculation,
\begin{align}
\underline{Z}^0+\sum_{N=0}^{N'}\left(\overline{Z}^N-\underline{Z}^N\right)-\underline{Z}^{N'+1}&\geq\sum_{N=0}^{N'}\int_N^{N+1}\left\{\overline{Z}(t)-\underline{Z}(t)\right\}dt\label{eqn:finitefinite8}\\
&\geq-\overline{Z}^0+\sum_{N=0}^{N'}\left(\overline{Z}^N-\underline{Z}^N\right)+\overline{Z}^{N'+1}.\label{eqn:finitefinite9}
\end{align}
From (\ref{eqn:finitefinite8}), (\ref{eqn:finitefinite9}) and Lemma \ref{lem:0iNleq12}, we have
\begin{align}
\dfrac{1}{2}+\sum_{N=0}^{N'}\left(\overline{Z}^N-\underline{Z}^N\right)\geq\int_0^{N'+1}\left\{\overline{Z}(t)-\underline{Z}(t)\right\}dt\geq-\dfrac{1}{2}+\sum_{N=0}^{N'}\left(\overline{Z}^N-\underline{Z}^N\right),
\end{align}
hence the assertion of the Lemma holds.\hfill$\blacksquare$
\bigskip
\begin{lemma}
\label{lem:evaluationofwi(t)}
There exists $L_2>0$ such that for $t>1$
\begin{align}
-\dfrac{\dot{z}_i(t)}{z_i(t)}-\dfrac{L_2}{(t-1)^2}<w_i(t)<-\dfrac{\dot{z}_i(t)}{z_i(t)},
\end{align}
where $\dot{z}_i(t)$ denotes the derivative of $z_i(t)$ by $t$.
\end{lemma}
\noindent{\bf Proof:}\ For $N<t<N+1$,
\begin{align}
\dot{z}_i(t)&={\rm the\ slope\ of\ }z_i(t)\\
&=\dfrac{z_i^{N+1}-z_i^N}{(N+1)-N}\\
&=z_i^{N+1}-z_i^N\\
&=-z_i^N\sum_{i'=1}^hq_{ii'}z_{i'}^N\label{eqn:slope}\\
&<-z_i(t)\sum_{i'=1}^hq_{ii'}z_{i'}(t)\ \ {\rm(by\ (\ref{eqn:zi(N)=ziNzi(N+1)=ziN+1})\ and\ Lemma\ \ref{lem:z_i(t)isstrictlydec})}\\
&=-z_i(t)w_i(t),
\end{align}
then, we have
\begin{align}
w_i(t)<-\dfrac{\dot{z}_i(t)}{z_i(t)}.\label{eqn:taui(t)leq}
\end{align}

Next, we bound $w_i(t)$ from below. Because $z_i(t)$ is decreasing, we have $z_i(t)>z_i^{N+1}$ for $N<t<N+1$. Therefore,
\begin{align}
z_i^N<z_i(t)+z_i^N-z_i^{N+1},\,i=1,\dots,h.\label{eqn:xiNleq}
\end{align}
Then, by (\ref{eqn:slope}) and (\ref{eqn:xiNleq}),
\begin{align}
\dot{z}_i(t)
&>-\left\{z_i(t)+z_i^N-z_i^{N+1}\right\}\sum_{i'=1}^hq_{ii'}\left\{z_{i'}(t)+z_{i'}^N-z_{i'}^{N+1}\right\}\\
&=-z_i(t)w_i(t)-\left[z_i(t)\sum_{i'=1}^hq_{ii'}\left(z_{i'}^N-z_{i'}^{N+1}\right)\right.\nonumber\\
&\ \ \ \ \left.+\left(z_i^N-z_i^{N+1}\right)\sum_{i'=1}^hq_{ii'}\left\{z_{i'}(t)+z_{i'}^N-z_{i'}^{N+1}\right\}\right].
\end{align}
hence,
\begin{align}
w_i(t)
&>-\dfrac{\dot{z}_i(t)}{z_i(t)}-\left[\sum_{i'=1}^hq_{ii'}\left(z_{i'}^N-z_{i'}^{N+1}\right)+\dfrac{z_i^N-z_i^{N+1}}{z_i(t)}\sum_{i'=1}^hq_{ii'}\left\{z_{i'}(t)+z_{i'}^N-z_{i'}^{N+1}\right\}\right]\\
&>-\dfrac{\dot{z}_i(t)}{z_i(t)}-\left[\sum_{i'=1}^hq_{ii'}\left(z_{i'}^N-z_{i'}^{N+1}\right)+\dfrac{z_i^N-z_i^{N+1}}{z_i^{N+1}}\sum_{i'=1}^hq_{ii'}\left\{z_{i'}(t)+z_{i'}^N-z_{i'}^{N+1}\right\}\right]\\
&=-\dfrac{\dot{z}_i(t)}{z_i(t)}-\left[\sum_{i'=1}^hq_{ii'}z_{i'}^Nw_{i'}^N+\dfrac{w_i^N}{1-w_i^N}\sum_{i'=1}^hq_{ii'}\left\{z_{i'}(t)+z_{i'}^Nw_{i'}^N\right\}\right]\\[2mm]
&>-\dfrac{\dot{z}_i(t)}{z_i(t)}-\left[\sum_{i'=1}^hq_{ii'}z_{i'}^Nw_{i'}^N+2w_i^N\sum_{i'=1}^hq_{ii'}\left\{z_{i'}(t)+z_{i'}^Nw_{i'}^N\right\}\right]\ \ \text{\rm (by\ Lemma\ \ref{lem:0<w_i^N<=1/2})}\\[2mm]
&>-\dfrac{\dot{z}_i(t)}{z_i(t)}-\dfrac{L_2}{N^2}\ \ (\,\exists L_2>0,{\rm \ by\ Lemma \ }\ref{lem:xiN<dfracL1N})\\[2mm]
&>-\dfrac{\dot{z}_i(t)}{z_i(t)}-\dfrac{L_2}{(t-1)^2}.\label{eqn:taui(t)geq}
\end{align}
\hfill$\blacksquare$
\subsection{The values of $t$ at which the slope of $\overline{Z}(t)$ or $\underline{Z}(t)$ changes}
Let $t_1,t_2,\dots,t_k,\dots$ be the values of $t$ at which the slope of $\overline{Z}(t)$ or $\underline{Z}(t)$ changes among all $0<t<\infty$, that is, $\dot{\overline{Z}}(t)$ or $\dot{\underline{Z}}(t)$ is discontinuous at $t_1,t_2,\dots,t_k,\dots$. See Fig. \ref{fig:changetime}. We put $t_0=0$ for convenience. Let $\cal T$ be the collection of $t_k$, i.e., ${\cal T}=\{t_k\}_{k=0,1,2,\dots}$.

When $t$ is a positive integer, both the slope of $\overline{Z}(t)$ and $\underline{Z}(t)$ change, so $t=1,2,\dots$ are included in $\cal T$. These values are shown in Fig. \ref{fig:changetime} by black squares. Furthermore, the values of $t$ at which $\overline{i}(t)$ that achieves $\overline{Z}(t)$ or $\underline{i}(t)$ that achieves $\underline{Z}(t)$ changes are also included in $\cal T$. These values are shown in Fig. \ref{fig:changetime} by black circles.

\bigskip
\bigskip
\bigskip

\begin{figure}[H]
\centering
\begin{overpic}[width=0.95\columnwidth]{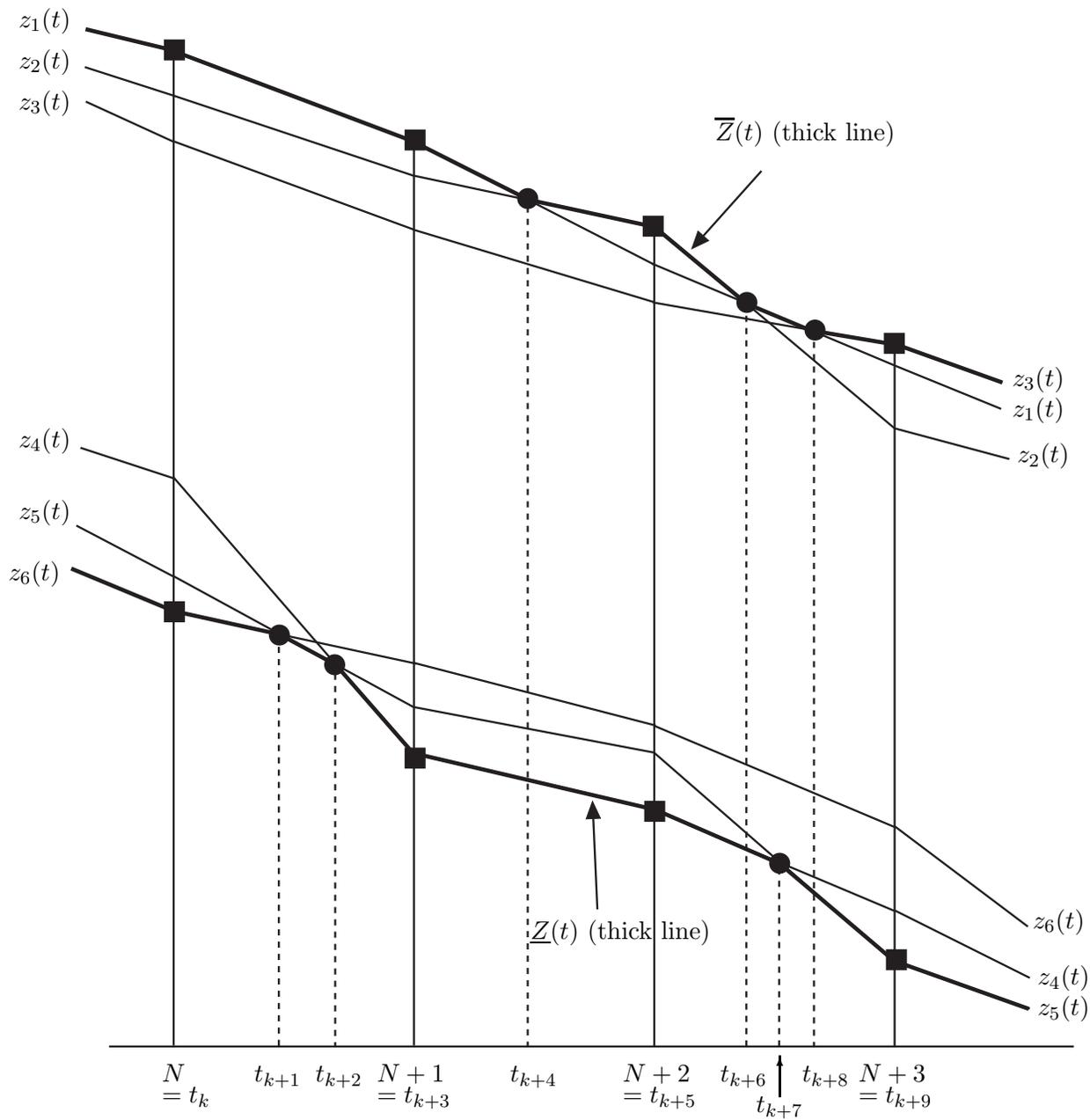}
\put(-5,100){$z_1(t)$}
\put(-5,96){$z_2(t)$}
\put(-5,92){$z_3(t)$}
\put(-5,59){$z_4(t)$}
\put(-5,52){$z_5(t)$}
\put(-6,46){$z_6(t)$}
\put(63,89){$\overline{Z}(t)$ (thick line)}
\put(45,11){$\underline{Z}(t)$ (thick line)}
\put(9,-3){$N$}
\put(9,-5){$=t_k$}
\put(30,-3){$N+1$}
\put(30,-5){$=t_{k+3}$}
\put(54,-3){$N+2$}
\put(54,-5){$=t_{k+5}$}
\put(77,-3){$N+3$}
\put(77,-5){$=t_{k+9}$}
\put(18,-3){$t_{k+1}$}
\put(24,-3){$t_{k+2}$}
\put(43,-3){$t_{k+4}$}
\put(63.5,-3){$t_{k+6}$}
\put(67,-6){$t_{k+7}$}
\put(69.3,-4.5){\vector(0,1){4}}
\put(71.5,-3){$t_{k+8}$}
\put(92,65){$z_3(t)$}
\put(92,62){$z_1(t)$}
\put(92.5,57.5){$z_2(t)$}
\put(94,12){$z_6(t)$}
\put(94.5,6.5){$z_4(t)$}
\put(94.5,3.5){$z_5(t)$}
\end{overpic}
\bigskip
\bigskip
\caption{The values of $t$ at which the slope of $\overline{Z}(t)$ or $\underline{Z}(t)$ changes}
\label{fig:changetime}
\end{figure}

\newpage

\begin{lemma}
\label{lem:overlineZt-underlineZtfinite}
\ $\ds\int_0^\infty\left\{\overline{Z}(t)-\underline{Z}(t)\right\}dt<\infty$.
\end{lemma}

\medskip

\noindent{\bf Proof:}\ By Lemma \ref{lem:overlineZt-underlineZt<L2tau-tau} and Lemma \ref{lem:evaluationofwi(t)},
\begin{align}
\overline{Z}(t)-\underline{Z}(t)
&\leq L_1\left\{w_{\,\overline{i}(t)}(t)-w_{\,\underline{i}(t)}(t)\right\}\\
&<L_1\left\{-\dfrac{\dot{z}_{\,\overline{i}(t)}(t)}{z_{\,\overline{i}(t)}(t)}+\dfrac{\dot{z}_{\,\underline{i}(t)}(t)}{z_{\,\underline{i}(t)}(t)}+\dfrac{L_2}{(t-1)^2}\right\},\ t\notin{\cal T}\\
&=L_1\left\{-\dfrac{{\dot{\overline{Z}}(t)}}{\,\overline{Z}(t)}+\dfrac{{\dot{\underline{Z}}(t)}}{\,\underline{Z}(t)}+\dfrac{L_2}{(t-1)^2}\right\},\ t\notin{\cal T}.\label{eqn:ZdotoverZ}
\end{align}
Because $2\in{\cal T}$, there exists an integer $k^\ast>1$ with $2=t_{k^\ast}$. Thus, by (\ref{eqn:ZdotoverZ}), for arbitrary integer $K>k^\ast$
\begin{align}
\sum_{k=0}^K&\int_{t_k}^{t_{k+1}}\left\{\overline{Z}(t)-\underline{Z}(t)\right\}dt\label{eqn:k=0toKintegral}\\
&=\sum_{k=0}^{k^\ast-1}\int_{t_k}^{t_{k+1}}\left\{\overline{Z}(t)-\underline{Z}(t)\right\}dt+\sum_{k=k^\ast}^K\int_{t_k}^{t_{k+1}}\left\{\overline{Z}(t)-\underline{Z}(t)\right\}dt\\
&<\int_0^{t_{k^\ast}}\left\{\overline{Z}(t)-\underline{Z}(t)\right\}dt+L_1\sum_{k=k^\ast}^K\left[-\log\overline{Z}(t)+\log\underline{Z}(t)-\dfrac{L_2}{t-1}\right]_{t_k}^{t_{k+1}}\\
&=\int_0^2\left\{\overline{Z}(t)-\underline{Z}(t)\right\}dt\nonumber\\
&\ \ \ \ \ \ \ +L_1\sum_{k=k^\ast}^K\left\{\log\dfrac{\overline{Z}(t_k)}{\underline{Z}(t_k)}-\log\dfrac{\overline{Z}(t_{k+1})}{\underline{Z}(t_{k+1})}+L_2\left(\dfrac{1}{t_k-1}-\dfrac{1}{t_{k+1}-1}\right)\right\}\\
&=\int_0^2\left\{\overline{Z}(t)-\underline{Z}(t)\right\}dt\nonumber\\
&\ \ \ \ \ \ \ +L_1\left\{\log\dfrac{\overline{Z}(t_{k^\ast})}{\underline{Z}(t_{k^\ast})}-\log\dfrac{\overline{Z}(t_{K+1})}{\underline{Z}(t_{K+1})}+L_2\left(\dfrac{1}{t_{k^\ast}-1}-\dfrac{1}{t_{K+1}-1}\right)\right\}\\[2mm]
&<\int_0^2\overline{Z}(t)dt+L_1\left\{\log\dfrac{\overline{Z}(2)}{\underline{Z}(2)}-\log\dfrac{\overline{Z}(t_{K+1})}{\underline{Z}(t_{K+1})}+L_2\left(\dfrac{1}{2-1}-\dfrac{1}{t_{K+1}-1}\right)\right\}\\[2mm]
&<2\overline{Z}(0)+L_1\left\{\log\dfrac{\overline{Z}(2)}{\underline{Z}(2)}+L_2\right\}.\label{eqn:upperbound}
\end{align}
The integral (\ref{eqn:k=0toKintegral}) has an upper bound (\ref{eqn:upperbound}) which does not depend on $K$, hence we obtain the Lemma.\hfill$\blacksquare$

\bigskip

\begin{lemma}
\label{eqn:sumofoverZN-underZNisfinite}
\ $\ds\sum_{N=0}^\infty\left(\overline{Z}^N-\underline{Z}^N\right)<\infty$.
\end{lemma}

\medskip

\noindent{\bf Proof:}\ The assertion holds by Lemma \ref{lem:finitefinite} and Lemma \ref{lem:overlineZt-underlineZtfinite}.\hfill$\blacksquare$

\bigskip

\begin{lemma}
\label{lem:overZN-underZNismonotonedecreasing}
\ The sequence $\left\{\overline{Z}^N-\underline{Z}^N\right\}_{N=0,1,\dots}$ is strictly decreasing.
\end{lemma}

\medskip

\noindent{\bf Proof:}\ We have
\begin{align}
\overline{Z}^{N+1}-\underline{Z}^{N+1}&=z_{\,\overline{i}(N+1)}^{N+1}-z_{\,\underline{i}(N+1)}^{N+1}\\
&=z_{\,\overline{i}(N+1)}^N-z_{\,\overline{i}(N+1)}^N\sum_{i'=1}^hq_{\,\overline{i}(N+1)i'}z_{i'}^N-z_{\,\underline{i}(N+1)}^N+z_{\,\underline{i}(N+1)}^N\sum_{i'=1}^hq_{\,\underline{i}(N+1)i'}z_{i'}^N\\
&<z_{\,\overline{i}(N+1)}^N-z_{\,\overline{i}(N+1)}^N\sum_{i'=1}^hq_{\,\overline{i}(N+1)i'}z_{\,\underline{i}(N)}^N-z_{\,\underline{i}(N+1)}^N+z_{\,\underline{i}(N+1)}^N\sum_{i'=1}^hq_{\,\underline{i}(N+1)i'}z_{\,\overline{i}(N)}^N\\
&=z_{\,\overline{i}(N+1)}^N\left(1-\underline{Z}^N\right)-z_{\,\underline{i}(N+1)}^N\left(1-\overline{Z}^N\right)\\
&\leq\overline{Z}^N\left(1-\underline{Z}^N\right)-\underline{Z}^N\left(1-\overline{Z}^N\right)\\
&=\overline{Z}^N-\underline{Z}^N.
\end{align}
\hfill$\blacksquare$

\bigskip

We will use the following theorem in our proof.

\medskip

\noindent{\bf Theorem C}\ (\cite{bro},\,{\it p}.31)\ \it Let $\{a_N\}_{N=0,1,\dots,}$ be a decreasing positive sequence. If $\sum_{N=0}^\infty a_N$ is convergent, then $Na_N\to0,\ N\to\infty$.\rm

\begin{lemma}
\label{lem:N(overZN-underZN)to0}
\ $\ds\lim_{N\to\infty}N\left(\overline{Z}^N-\underline{Z}^N\right)=0$.
\end{lemma}

\medskip

\noindent{\bf Proof:}\ The assertion holds by Lemma \ref{eqn:sumofoverZN-underZNisfinite}, Lemma \ref{lem:overZN-underZNismonotonedecreasing} and Theorem C.\hfill$\blacksquare$

\bigskip

\setcounter{theorem}{3}

\begin{theorem}
\ $\ds\lim_{N\to\infty}N\overline{Z}^N=1,\ \ds\lim_{N\to\infty}N\underline{Z}^N=1$.
\end{theorem}

\medskip

\noindent{\bf Proof:}\ By Lemma \ref{lem:limsupNtoinftyNunderlineZNleq1} and Lemma \ref{lem:N(overZN-underZN)to0}, we have
\begin{align}
\limsup_{N\to\infty}N\overline{Z}^N&=\limsup_{N\to\infty}\left(N\overline{Z}^N-N\underline{Z}^N+N\underline{Z}^N\right)\\
&\leq\limsup_{N\to\infty}N\left(\overline{Z}^N-\underline{Z}^N\right)+\limsup_{N\to\infty}N\underline{Z}^N\\
&\leq1,
\end{align}
thus, together with Lemma \ref{lem:liminfNtoinftyNoverlineZNgeq1}, we have
\begin{align}
\lim_{N\to\infty}N\overline{Z}^N=1.\label{eqn:limNtoinftyNxi1N=1}
\end{align}
Further, we have
\begin{align}
\lim_{N\to\infty}N\underline{Z}^N&=\lim_{N\to\infty}\left(N\underline{Z}^N-N\overline{Z}^N+N\overline{Z}^N\right)\\
&=-\lim_{N\to\infty}N(\overline{Z}^N-\underline{Z}^N)+\lim_{N\to\infty}N\overline{Z}^N\\
&=1.\label{eqn:limNtoinftyNxi3N=1}
\end{align}
\hfill$\blacksquare$

As stated above, Theorem \ref{the:hasamiuchi} implies Theorem \ref{the:main}, the goal of this study.

\section{Numerical examples}
Based on Example 7 in \cite{nak5}, consider the following channel matrix $\Phi$:

\begin{align}
\label{eqn:examplePhi}
\Phi=\begin{pmatrix}P^1\\ P^2\\ P^3\\ P^4\\ P^5\end{pmatrix}
=\begin{pmatrix}
0.6 & 0.1 & 0.1 & 0.1 & 0.1\\
0.1 & 0.6 & 0.1 & 0.1 & 0.1\\
P^3_1 & P^3_2 & P^3_3 & P^3_4 & P^3_5\\
P^4_1 & P^4_2 & P^4_3 & P^4_4 & P^4_5\\
P^5_1 & P^5_2 & P^5_3 & P^5_4 & P^5_5
\end{pmatrix},
\end{align}
where $P^3,\ P^4$ and $P^5$ will be fixed later.

Let the output distribution $Q^\ast$ be the midpoint of $P^1$ and $P^2$, i.e.,
\begin{align}
Q^\ast=(0.35,\,0.35,\,0.1,\,0.1,\,0.1),
\end{align}
and define a real number $C$ as
\begin{align}
D(P^1\|Q^\ast)=D(P^2\|Q^\ast)=0.198121603\equiv C.
\end{align}

We will choose $P^3,\,P^4,\,P^5$ so that the above $C$ is the channel capacity of $\Phi$ and $Q^\ast$ is the capacity achieving output distribution and $\{3,\,4,\,5\}$ are type-II indices. That is, we will choose $P^3,\,P^4,\,P^5$ that satisfy the following conditions:
\begin{align}
&D(P^i\|Q^\ast)=C,\ i=3,4,5,\label{eqn:sphere_eq}\\
&P^i_j\geq0,\ j=1,\dots,5,\ i=3,4,5,\ \sum_{j=1}^5P^i_j=1,\ i=3,4,5,\label{eqn:sum=1}\\
&{\mbox{\rm rank}}\,\Phi=5.\label{eqn:rank=5}
\end{align}

See Fig. \ref{fig:circle}.

\begin{figure}[H]
\begin{center}
\begin{overpic}[width=0.65\columnwidth]{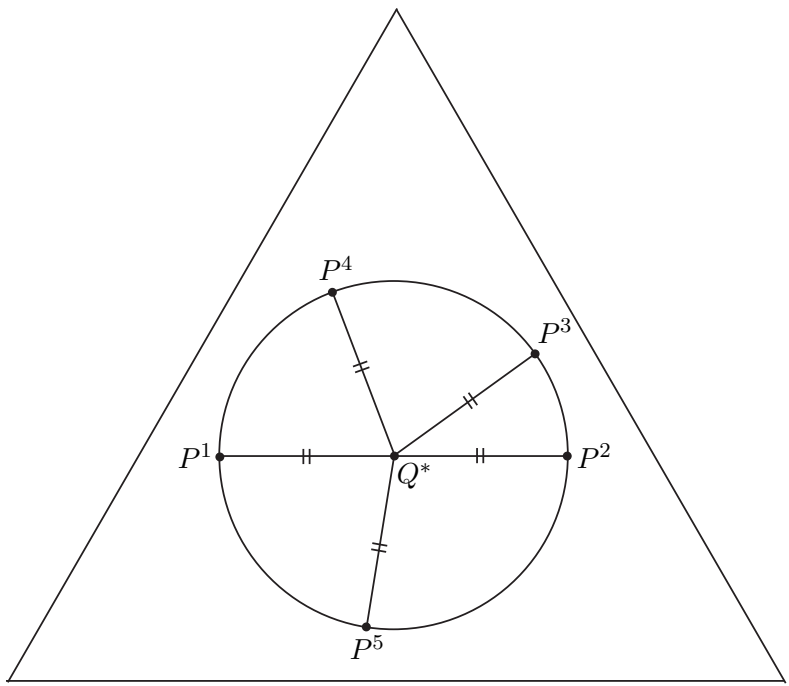}
\put(22,28){$P^1$}
\put(73,28){$P^2$}
\put(68,44){$P^3$}
\put(40,52){$P^4$}
\put(44,3.5){$P^5$}
\put(50,26){$Q^\ast$}
\end{overpic}
\end{center}
\caption{How to choose $P^3,\,P^4,\,P^5$ that satisfy (\ref{eqn:sphere_eq}), (\ref{eqn:sum=1}), (\ref{eqn:rank=5}).}
\label{fig:circle}
\end{figure}

The channel matrix $\Phi$ of (\ref{eqn:examplePhi}) is constructed using three solutions $P^3,P^4,P^5$ that satisfy all the conditions (\ref{eqn:sphere_eq}), (\ref{eqn:sum=1}), (\ref{eqn:rank=5}). In this case, $C$ is the channel capacity of $\Phi$ and $Q^\ast$ is the capacity achieving output distribution. The capacity achieving input distribution $\bm\lambda^\ast$ is
\begin{align}
\bm\lambda^\ast=(0.5,\,0.5,\,0,\,0,\,0).\label{eqn:lambdaasterisk}
\end{align}

Let us find the recurrence formula (\ref{eqn:1/Norderzenkashiki}) for $\Phi$ of (\ref{eqn:examplePhi}) obtained in this way. That is, find the matrix $R=(r_{ii'})_{i,i'=3,4,5}$.
\subsection{Calculation of Jacobian matrix $J(\bm\lambda^\ast)$}
For $\Phi$ of (\ref{eqn:examplePhi}), we see from (\ref{eqn:sphere_eq}), (\ref{eqn:lambdaasterisk}), that the type-I indices are ${\cal I}_{\rm I}=\{1,2\}$ and type-II indices are ${\cal I}_{\rm II}=\{3,4,5\}$, and there are no type-III indices. Therefore, (\ref{eqn:J1AJ2}) and (\ref{eqn:JJ'JII}) are the same and they equal
\begin{align}
\label{eqn:J1J2noJ3}
J(\bm\lambda^\ast)=\begin{pmatrix}
\,J^{\rm I} & O \,\\[1mm]
\,\ast & J^{\rm II}
\end{pmatrix}.
\end{align}
From Theorem 1 in \cite{nak5}, we have
\begin{align}
J(\bm\lambda^\ast)
&=\begin{pmatrix}
\,1+\lambda_1^\ast D_{1,1}^\ast & \lambda_2^\ast D_{1,2}^\ast &0&0& 0\,\\
\,\lambda_1^\ast D_{2,1}^\ast & 1+\lambda_2^\ast D_{2,2}^\ast &0&0& 0 \\
\,\lambda_1^\ast D_{3,1}^\ast & \lambda_2^\ast D_{3,2}^\ast & 1&0&0\\
\,\lambda_1^\ast D_{4,1}^\ast & \lambda_2^\ast D_{4,2}^\ast & 0&1&0\\
\,\lambda_1^\ast D_{5,1}^\ast & \lambda_2^\ast D_{5,2}^\ast & 0&0&1
\end{pmatrix},\label{eqn:Jacobi_in_example}
\end{align}
where
\begin{align}
D^\ast_{i,i'}=-\ds\sum_{j=1}^5\dfrac{P^i_jP^{i'}_j}{Q^\ast_j},\ i,i'=1,\dots,5.\label{eqn:def_of_Dii'}
\end{align}
See Lemma 4 in \cite{nak5}.

Let us compute $D^\ast_{i,i'}$ of (\ref{eqn:def_of_Dii'}) for $\Phi$ of (\ref{eqn:examplePhi}). Indeed, we have
\begin{align}
D^\ast_{1,1}=-\dfrac{19}{14},\ D^\ast_{1,2}=D^\ast_{2,1}=-\dfrac{9}{14},\ D^\ast_{2,2}=-\dfrac{19}{14},\label{eqn:D11}
\end{align}
and further, for $i=3,4,5$,
\begin{align}
D^\ast_{1,i}=D^\ast_{i,1}=-1-\dfrac{5}{7}P^i_1+\dfrac{5}{7}P^i_2,\ \ \ D^\ast_{2,i}=D^\ast_{i,2}=-1+\dfrac{5}{7}P^i_1-\dfrac{5}{7}P^i_2.\label{eqn:D1i}
\end{align}
Therefore, by substituting (\ref{eqn:D11}) and (\ref{eqn:D1i}) into (\ref{eqn:Jacobi_in_example}), we have
\begin{align}
J(\bm\lambda^\ast)
&=\begin{pmatrix}
\,\dfrac{9}{28} & -\dfrac{9}{28} &0&0& 0\,\\[4mm]
\,-\dfrac{9}{28} & \dfrac{9}{28} &0&0& 0 \\[4mm]
\,-\dfrac{1}{2}-\dfrac{5}{14}P^3_1+\dfrac{5}{14}P^3_2 & -\dfrac{1}{2}+\dfrac{5}{14}P^3_1-\dfrac{5}{14}P^3_2& 1&0&0\\[4mm]
\,-\dfrac{1}{2}-\dfrac{5}{14}P^4_1+\dfrac{5}{14}P^4_2 & -\dfrac{1}{2}+\dfrac{5}{14}P^4_1-\dfrac{5}{14}P^4_2& 0&1&0\\[4mm]
\,-\dfrac{1}{2}-\dfrac{5}{14}P^5_1+\dfrac{5}{14}P^5_2 & -\dfrac{1}{2}+\dfrac{5}{14}P^5_1-\dfrac{5}{14}P^5_2& 0&0&1
\end{pmatrix}.\label{eqn:Jacobi_in_example_gutai}
\end{align}

Furthermore, for $i,i'=3,4,5$, we have
\begin{align}
D^\ast_{i,i'}=D^\ast_{i',i}=-\left\{\dfrac{20}{7}P^i_1P^{i'}_1+\dfrac{20}{7}P^i_2P^{i'}_2+10\left(P^i_3P^{i'}_3+P^i_4P^{i'}_4+P^i_5P^{i'}_5\right)\right\}.
\end{align}
\subsection{Eigenvalues and eigenvectors of $J(\bm\lambda^\ast)$}
From (\ref{eqn:Jacobi_in_example_gutai}), the eigenvalues of $J(\bm\lambda^\ast)$ are 
\begin{align}
(\theta_1,\,\theta_2,\,\theta_3,\,\theta_4,\,\theta_5)=(0,\,9/14,\,1,\,1,\,1).
\end{align}
Taking appropriate eigenvector $\bm{a}_i$ for each eigenvalue $\theta_i$, then by (\ref{eqn:a1...am})-(\ref{eqn:A1A2definition}), we have
\begin{align}
A=\begin{pmatrix}
\,1 & 1 & 0 & 0 & 0\,\\
\,1 & -1 & 0 & 0 & 0\,\\
\,1 & 2P^3_1-2P^3_2 & 1 & 0 & 0\,\\
\,1 & 2P^4_1-2P^4_2 & 0 & 1 & 0\,\\
\,1 & 2P^5_1-2P^5_2 & 0 & 0 & 1\,
\end{pmatrix},\ 
\end{align}
\begin{align}
A_1=\begin{pmatrix}
\,1 & 1\,\\
\,1 & -1\,
\end{pmatrix},
\end{align}
\begin{align}
A_2=\begin{pmatrix}
\,1 & 2P^3_1-2P^3_2 \,\\
\,1 & 2P^4_1-2P^4_2 \,\\
\,1 & 2P^5_1-2P^5_2 \,
\end{pmatrix}.
\end{align}
Then, from $\bm\mu^N_{\rm I,III}=-\bm\mu^N_{\rm II}A_2A_1^{-1}$ of (\ref{eqn:bmmuN{I,III}A1+bmmuN{II}A2=bm0}), we have
\begin{align}
\left(\mu_1^N,\mu_2^N\right)=-\dfrac{1}{2}\left(\mu_3^N,\mu_4^N,\mu_5^N\right)\begin{pmatrix}
\,1 & 2P^3_1-2P^3_2 \,\\
\,1 & 2P^4_1-2P^4_2 \,\\
\,1 & 2P^5_1-2P^5_2 \,
\end{pmatrix}
\begin{pmatrix}
\,1 & 1\,\\
\,1 & -1\,
\end{pmatrix},
\end{align}
thus,
\begin{align}
\mu^N_1&=\left(-\dfrac{1}{2}-P^3_1+P^3_2\right)\mu_3^N+\left(-\dfrac{1}{2}-P^4_1+P^4_2\right)\mu_4^N+\left(-\dfrac{1}{2}-P^5_1+P^5_2\right)\mu_5^N,\label{eqn:mu1bymu345}\\[2mm]
\mu^N_2&=\left(-\dfrac{1}{2}+P^3_1-P^3_2\right)\mu_3^N+\left(-\dfrac{1}{2}+P^4_1-P^4_2\right)\mu_4^N+\left(-\dfrac{1}{2}+P^5_1-P^5_2\right)\mu_5^N.\label{eqn:mu2bymu345}
\end{align}
\subsection{On the Hessian matrix $H_i(\bm\lambda^\ast),\,i=3,4,5$}
For $i=3,4,5$, we will calculate the Hessian matrices $H_3(\bm\lambda^\ast),\,H_4(\bm\lambda^\ast),\,H_5(\bm\lambda^\ast)$. From Theorem 6 in \cite{nak5}, we have
\begin{align}
H_3(\bm\lambda^\ast)=\begin{pmatrix}
\,0 & 0 & D^\ast_{1,3} & 0 & 0\,\\
\,0 & 0 & D^\ast_{2,3} & 0 & 0\,\\
\,D^\ast_{1,3} & D^\ast_{2,3} & 2D^\ast_{3,3} & D^\ast_{3,4} & D^\ast_{3,5}\,\\
\,0 & 0 & D^\ast_{3,4} & 0 & 0\,\\
\,0 & 0 & D^\ast_{3,5} & 0 & 0\,
\end{pmatrix},\label{eqn:H3}
\end{align}
\begin{align}
H_4(\bm\lambda^\ast)=\begin{pmatrix}
\,0 & 0 & 0 & D^\ast_{1,4} & 0\,\\
\,0 & 0 & 0 & D^\ast_{2,4} & 0\,\\
\,0 & 0 & 0 & D^\ast_{3,4} & 0\,\\
\,D^\ast_{1,4} & D^\ast_{2,4} & D^\ast_{3,4} & 2D^\ast_{4,4} & D^\ast_{4,5}\,\\
\,0 & 0 & 0 & D^\ast_{4,5} & 0\,
\end{pmatrix},\label{eqn:H4}
\end{align}
\begin{align}
H_5(\bm\lambda^\ast)=\begin{pmatrix}
\,0 & 0 & 0 & 0 & D^\ast_{1,5}\,\\
\,0 & 0 & 0 & 0 & D^\ast_{2,5}\,\\
\,0 & 0 & 0 & 0 & D^\ast_{3,5}\,\\
\,0 & 0 & 0 & 0 & D^\ast_{4,5}\,\\
\,D^\ast_{1,5} & D^\ast_{2,5} & D^\ast_{3,5} & D^\ast_{4,5} & 2D^\ast_{5,5}\,
\end{pmatrix}.\label{eqn:H5}
\end{align}

Then, for $\bm\mu^N=\left(\mu_1^N,\mu_2^N,\mu_3^N,\mu_4^N,\mu_5^N\right)$, we will calculate $(1/2)\bm\mu^NH_i(\bm\lambda^\ast)\,^t\!\bm\mu^N,\,i=3,4,5$.

First, for $i=3$, from (\ref{eqn:mu1bymu345}), (\ref{eqn:mu2bymu345}), (\ref{eqn:H3}), we have
\begin{align}
\dfrac{1}{2}\bm\mu^NH_3(\bm\lambda^\ast)\,^t\!\bm\mu^N=\mu_3^N\left(r_{33}\mu_3^N+r_{34}\mu_4^N+r_{35}\mu_5^N\right),
\end{align}
where
\begin{align}
r_{33}&=1+\dfrac{10}{7}\left(P^3_1-P^3_2\right)^2+D^\ast_{3,3},\\[2mm]
r_{34}&=1+\dfrac{10}{7}\left(P^3_1-P^3_2\right)\left(P^4_1-P^4_2\right)+D^\ast_{3,4},\\[2mm]
r_{35}&=1+\dfrac{10}{7}\left(P^3_1-P^3_2\right)\left(P^5_1-P^5_2\right)+D^\ast_{3,5}.
\end{align}

Similarly, for $i=4$, from (\ref{eqn:mu1bymu345}), (\ref{eqn:mu2bymu345}), (\ref{eqn:H4}), we have
\begin{align}
\dfrac{1}{2}\bm\mu^NH_4(\bm\lambda^\ast)\,^t\!\bm\mu^N=\mu_4^N\left(r_{43}\mu_3^N+r_{44}\mu_4^N+r_{45}\mu_5^N\right),
\end{align}
where
\begin{align}
r_{43}&=1+\dfrac{10}{7}\left(P^4_1-P^4_2\right)\left(P^3_1-P^3_2\right)+D^\ast_{3,4},\\[2mm]
r_{44}&=1+\dfrac{10}{7}\left(P^4_1-P^4_2\right)^2+D^\ast_{4,4},\\[2mm]
r_{45}&=1+\dfrac{10}{7}\left(P^4_1-P^4_2\right)\left(P^5_1-P^5_2\right)+D^\ast_{4,5}.
\end{align}

For $i=5$, for $i=4$, from (\ref{eqn:mu1bymu345}), (\ref{eqn:mu2bymu345}), (\ref{eqn:H5}), we have
\begin{align}
\dfrac{1}{2}\bm\mu^NH_5(\bm\lambda^\ast)\,^t\!\bm\mu^N=\mu_5^N\left(r_{53}\mu_3^N+r_{54}\mu_4^N+r_{55}\mu_5^N\right),
\end{align}
where
\begin{align}
r_{53}&=1+\dfrac{10}{7}\left(P^5_1-P^5_2\right)\left(P^3_1-P^3_2\right)+D^\ast_{3,5},\\[2mm]
r_{54}&=1+\dfrac{10}{7}\left(P^5_1-P^5_2\right)\left(P^4_1-P^4_2\right)+D^\ast_{4,5},\\[2mm]
r_{55}&=1+\dfrac{10}{7}\left(P^5_1-P^5_2\right)^2+D^\ast_{5,5}.
\end{align}

Then, for
\begin{align}
R=\begin{pmatrix}
r_{33} & r_{34} & r_{35}\\
r_{43} & r_{44} & r_{45}\\
r_{53} & r_{54} & r_{55}
\end{pmatrix},\label{eqn:Rmatrix_components}
\end{align}
we will solve the equation
\begin{align}
\bm\sigma=-\bm1(^tR)^{-1},\ \bm\sigma=(\sigma_3,\sigma_4,\sigma_5),
\end{align}
in the assumption (iii). If $R$ of (\ref{eqn:Rmatrix_components}) satisfies the assumptions (i)-(iv), we have, from Theorem \ref{the:main},
\bigskip

$\ds\lim_{N\to\infty}N\mu_i^N$\\
\begin{empheq}[left={\empheqlbrace}]{alignat=1}
&\left(-\dfrac{1}{2}-P^3_1+P^3_2\right)\sigma_3+\left(-\dfrac{1}{2}-P^4_1+P^4_2\right)\sigma_4+\left(-\dfrac{1}{2}-P^5_1+P^5_2\right)\sigma_5,\ i=1,\\
&\left(-\dfrac{1}{2}+P^3_1-P^3_2\right)\sigma_3+\left(-\dfrac{1}{2}+P^4_1-P^4_2\right)\sigma_4+\left(-\dfrac{1}{2}+P^5_1-P^5_2\right)\sigma_5,\ i=2,\\
&\ \ \sigma_i,\ i=3,4,5.
\end{empheq}
\subsection{How to make a channel matrix of (\ref{eqn:examplePhi})}
We will show here how to make a channel matrix of (\ref{eqn:examplePhi}), i.e., how to choose probability distributions $P^3,\ P^4$ and $P^5$.

As ${\cal Y}=\{y_1,\dots,y_5\}$, take a probability distribution $P$ from $\Delta({\cal Y})$ whose components are non-negative integral multiples of 1/3. There are 35 such probability distributions $P$ in total. Consider a half-line $P_s=(1-s)Q^\ast+sP,\ s\geq0$ starting from $Q^\ast$ and passing through $P$. Determine $s$ by a numerical search so that $D(P_s||Q^\ast)=C$. Let $P^3\equiv P_s$ be a probability distribution determined in this way. Similarly, 35 $P^3$ are obtained. In the same way, 35 $P^4$ and 35 $P^5$ are obtained, then, $_{35}C_3=6545$ combinations of $P^3,\,P^4,\,P^5$ are obtained because we choose so that $P^3\neq P^4\neq P^5\neq P^3$. Then, using those $P^3,\,P^4,\,P^5$, we obtain 6545 channel matrices $\Phi$ of (\ref{eqn:examplePhi}).

Examining these 6545 $\Phi$, we found that 109 of them satisfy the assumptions (i)-(iv). This is $109/6545\doteqdot1.7\%$ of the total. We had expected that $100\%$ of the total would satisfy (i)-(iv), but in fact, it is a very small percentage.

We have already proved in Chapter \ref{sec:shuusokushoumei} that the Arimoto-Blahut algorithm converges at the speed of the order $O(1/N)$ for those 109 $\Phi$ satisfying (i)-(iv), but in that proof, we used all the assumptions (i)-(iv). Thus, the present proof cannot be used for $\Phi$ if one or more among (i)-(iv) are not satisfied. Nonetheless, it seems correct that for all $\Phi$ the speed of convergence is the order $O(1/N)$ by a lot of numerical experiments. So, we would like to prove it somehow, but we have no clue at this moment. This is an issue for the future.

\newpage
\subsection{Channel matrices that satisfy the assumptions (i)-(iv)}
First, we will compare the numerical values $N\mu_i^N$ with the theoretical value $\sigma_i$ obtained in this paper for the following two channel matrices $\Phi^{(1)},\,\Phi^{(2)}$ among 109 $\Phi$ that satisfy (i)-(iv).
\begin{example}
\label{exa:example1}
\rm 
We will show below the channel matrix $\Phi^{(1)}$ together with $R^{(1)}$ and $\bm\sigma^{(1)}$ for $\Phi^{(1)}$, where $R^{(1)}$ and $\bm\sigma^{(1)}$ are $R$ and $\bm\sigma$ defined in section \ref{sec:canonicalform}, respectively:
\begin{align}
\Phi^{(1)}&=\begin{pmatrix}
0.600 & 0.100 & 0.100 & 0.100 & 0.100\\
0.100 & 0.600 & 0.100 & 0.100 & 0.100\\
0.231 & 0.231 & 0.066 & 0.179 & 0.292\\
0.161 & 0.341 & 0.226 & 0.226 & 0.046\\
0.341 & 0.161 & 0.226 & 0.046 & 0.226
\end{pmatrix},\\
R^{(1)}&=\begin{pmatrix}\label{eqn:R1}
-0.522 & -0.020 & -0.223\\
-0.020 & -0.401 & -0.078\\
-0.223 & -0.078 & -0.401
\end{pmatrix},\\
\bm\sigma^{(1)}&=(-2.251,\ -2.532,\ 1.241,\ 2.161,\ 1.381).\label{eqn:sigma1}
\end{align}
We set the initial distribution as $\bm\lambda^0=(0.200,\,0.200,\,0.200,\,0.200,\,0.200)$.

From (\ref{eqn:R1}), (\ref{eqn:sigma1}), we see that all the assumptions (i)-(iv) are satisfied.

Fig. \ref{fig:1} shows the values of $N\mu_i^N$ and $\sigma^{(1)}_i$ of (\ref{eqn:sigma1}). From the Fig. \ref{fig:1}, $N\mu_i^N$ converges to a value close enough to $\sigma^{(1)}_i$ at $N=1000$.
\begin{figure}[H]
\begin{center}
\begin{overpic}[width=0.7\columnwidth]{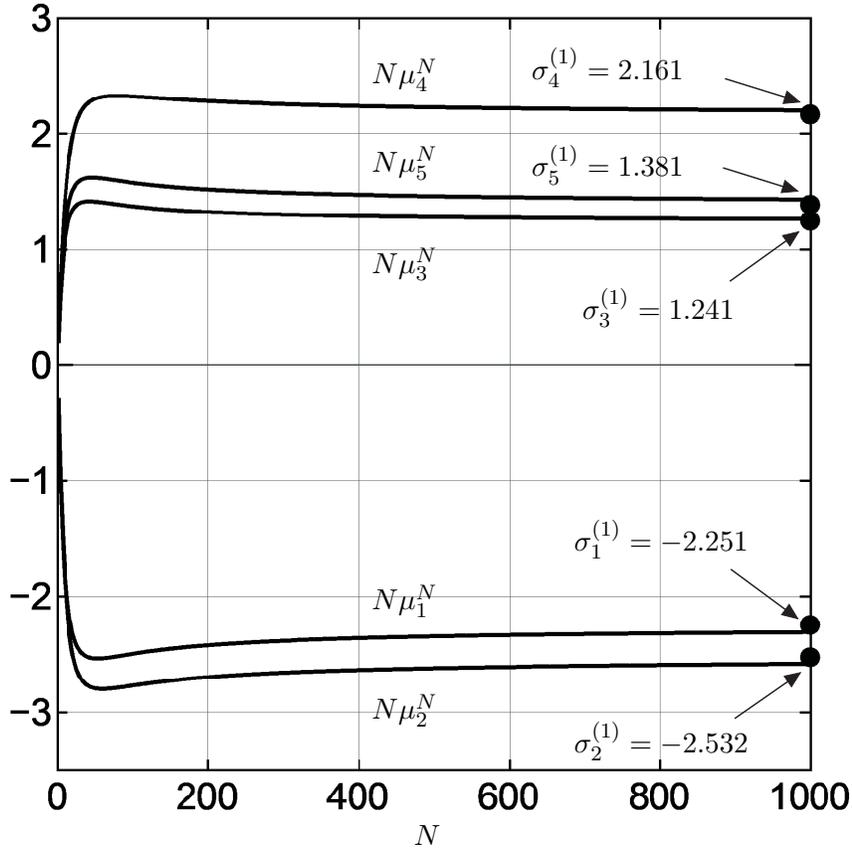}
\put(43,25){$N\mu^N_1$}
\put(43,12){$N\mu^N_2$}
\put(43,65){$N\mu^N_3$}
\put(43,77){$N\mu^N_5$}
\put(43,87.5){$N\mu^N_4$}
\put(67,8){$\sigma^{(1)}_2=-2.532$}
\put(67,32){$\sigma^{(1)}_1=-2.251$}
\put(68,59.5){$\sigma^{(1)}_3=1.241$}
\put(62,76.5){$\sigma^{(1)}_5=1.381$}
\put(62,88){$\sigma^{(1)}_4=2.161$}
\put(48,-3){$N$}
\end{overpic}
\end{center}
\caption{Comparison of $N\mu^N_i$ and $\sigma^{(1)}_i$ of Example \ref{exa:example1}.}
\label{fig:1}
\end{figure}
\end{example}
\newpage
\begin{example}
\label{exa:example2}
\rm 
We will show the channel matrix $\Phi^{(2)}$ together with $R^{(2)}$ and $\bm\sigma^{(2)}$ for $\Phi^{(2)}$:
\begin{align}
\Phi^{(2)}&=\begin{pmatrix}
0.600 & 0.100 & 0.100 & 0.100 & 0.100\\
0.100 & 0.600 & 0.100 & 0.100 & 0.100\\
0.231 & 0.231 & 0.066 & 0.179 & 0.292\\
0.161 & 0.341 & 0.226 & 0.226 &	0.046\\
0.522 & 0.160 & 0.046 & 0.227 & 0.046
\end{pmatrix},\\
R^{(2)}&=\begin{pmatrix}\label{eqn:R2}
-0.522 & -0.020 & -0.020\\
-0.020 & -0.401 & -0.126\\
-0.020 & -0.126 & -0.221
\end{pmatrix},\\
\bm\sigma^{(2)}&=(-4.415,\ -2.221,\ 1.723,\ 1.263,\ 3.650).\label{eqn:sigma2}
\end{align}
We set the initial distribution as $\bm\lambda^0=(0.200,\,0.200,\,0.200,\,0.200,\,0.200)$.

From (\ref{eqn:R2}), (\ref{eqn:sigma2}), we see that all the assumptions (i)-(iv) are satisfied.

Fig. \ref{fig:2} shows the values of $N\mu_i^N$ and $\sigma^{(2)}_i$ of (\ref{eqn:sigma2}). From the Fig. \ref{fig:2}, $N\mu_i^N$ converges to a value close enough to $\sigma^{(2)}_i$ at $N=1000$.
\begin{figure}[H]
\begin{center}
\begin{overpic}[width=0.7\columnwidth]{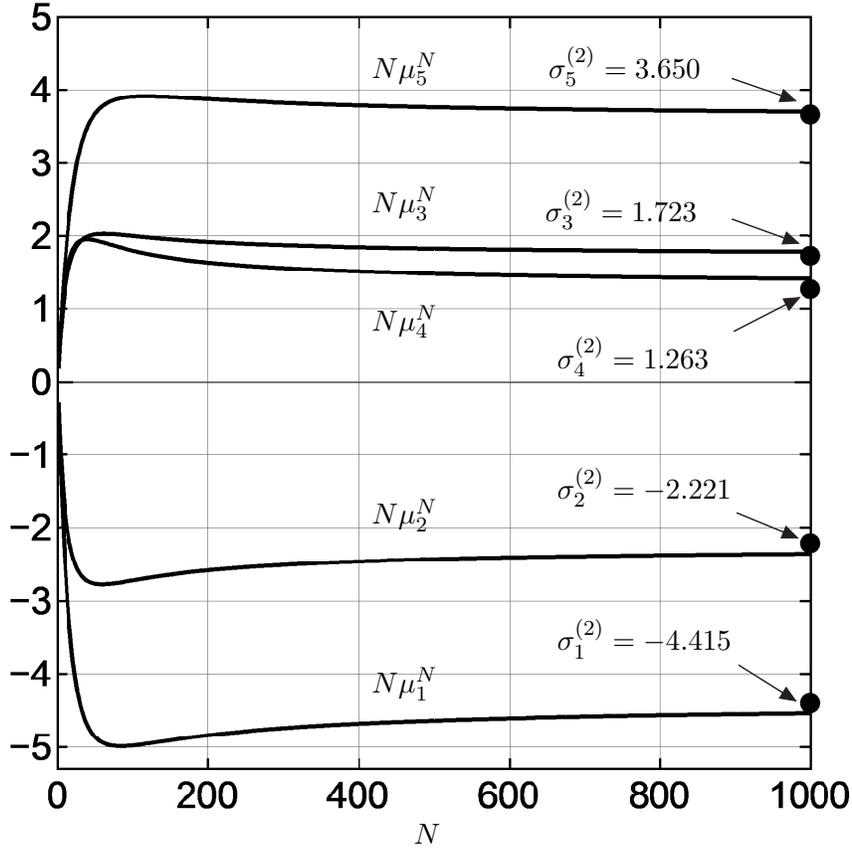}
\put(43,15){$N\mu_1^N$}
\put(43,35){$N\mu_2^N$}
\put(43,58){$N\mu_4^N$}
\put(43,72){$N\mu_3^N$}
\put(43,88){$N\mu_5^N$}
\put(65,20){$\sigma^{(2)}_1=-4.415$}
\put(65,38){$\sigma^{(2)}_2=-2.221$}
\put(65,53.5){$\sigma^{(2)}_4=1.263$}
\put(63.6,71){$\sigma^{(2)}_3=1.723$}
\put(64,88){$\sigma^{(2)}_5=3.650$}
\put(48,-3){$N$}
\end{overpic}
\end{center}
\caption{Comparison of $N\mu^N_i$ and $\sigma^{(2)}_i$ of Example \ref{exa:example2}.}
\label{fig:2}
\end{figure}
\end{example}
\newpage
\subsubsection{Channel matrices that do not satisfy the assumptions (i)-(iv)}
Next, we will show the numerical values $N\mu_i^N$ for the following two channel matrices $\Phi^{(3)},\,\Phi^{(4)}$ among $6436(=6545-109)$ $\Phi$ that do not satisfy (i)-(iv).
\begin{example}
\rm 
We will show the channel matrix $\Phi^{(3)}$ together with $R^{(3)}$ and $\bm\sigma^{(3)}$ for $\Phi^{(3)}$:
\begin{align}
\Phi^{(3)}&=\begin{pmatrix}
0.600 & 0.100 & 0.100 & 0.100 & 0.100\\
0.100 & 0.600 & 0.100 & 0.100 & 0.100\\
0.260 & 0.260 & 0.074 & 0.074 & 0.331\\
0.231 & 0.231 & 0.066 & 0.179 & 0.292\\
0.215 & 0.344 & 0.061 & 0.319 & 0.061
\end{pmatrix},\\
R^{(3)}&=\begin{pmatrix}\label{eqn:R3}
-0.594 & -0.493 & 0.099\\
-0.493 & -0.522 & -0.160\\
0.099 & -0.160 & -0.536
\end{pmatrix},\\
\bm\sigma^{(3)}&=(-75.23,\ -145.2,\ 528.0,\ -579.6,\ 272.1).\label{eqn:sigma3}
\end{align}
We set the initial distribution as $\bm\lambda^0=(0.200,\,0.200,\,0.200,\,0.200,\,0.200)$.

From (\ref{eqn:R3}), (\ref{eqn:sigma3}), we see that the assumptions (i), (iii), (iv) are not satisfied.

Fig. \ref{fig:3} shows the values of $N\mu_i^N$. From the Fig. \ref{fig:3}, $N\mu_i^N$ looks to converge to a constant value, but they are completely different from the values $\sigma^{(3)}_i$ of (\ref{eqn:sigma3}).
\begin{figure}[H]
\begin{center}
\begin{overpic}[width=0.7\columnwidth]{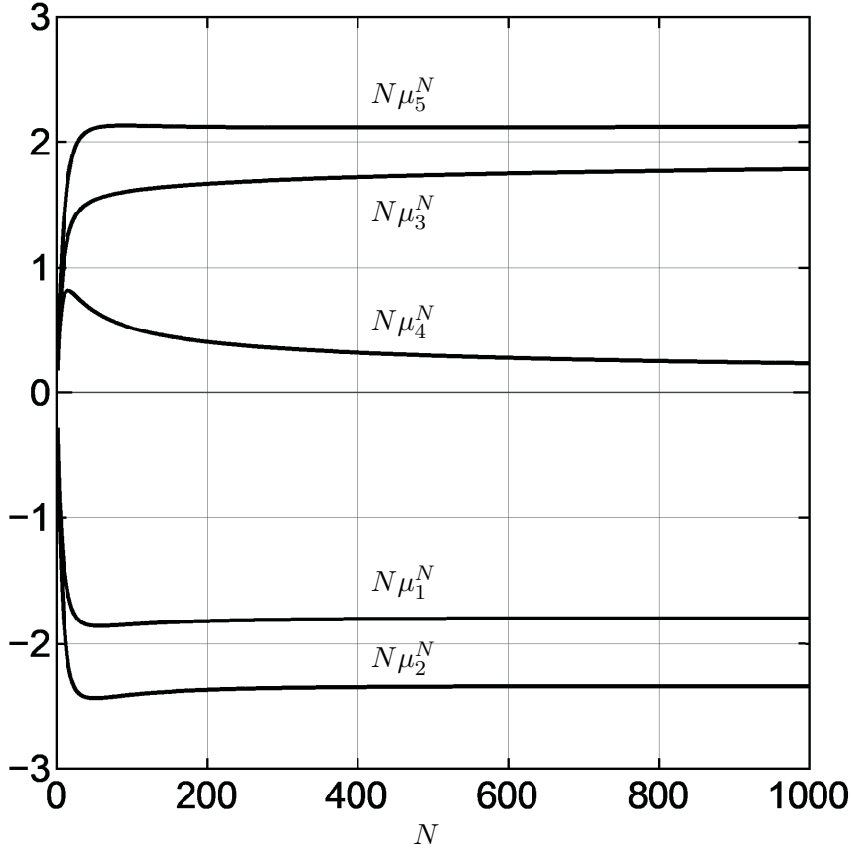}
\put(43,27){$N\mu_1^N$}
\put(43,18){$N\mu_2^N$}
\put(43,58){$N\mu_4^N$}
\put(43,71){$N\mu_3^N$}
\put(43,85){$N\mu_5^N$}
\put(48,-3){$N$}
\end{overpic}
\end{center}
\caption{$N\mu_i^N$ of Example 3.}
\label{fig:3}
\end{figure}
\end{example}
\newpage
\begin{example}
\rm 
We will show the channel matrix $\Phi^{(4)}$ together with $R^{(4)}$ and $\bm\sigma^{(4)}$ for $\Phi^{(4)}$:
\begin{align}
\Phi^{(4)}&=\begin{pmatrix}
0.600 & 0.100 & 0.100 & 0.100 & 0.100\\
0.100 & 0.600 & 0.100 & 0.100 & 0.100\\
0.522 & 0.160 & 0.046 & 0.046 & 0.227\\
0.522 & 0.160 & 0.046 & 0.227 & 0.046\\
0.522 & 0.160 & 0.227 & 0.046 & 0.046
\end{pmatrix},\\
R^{(4)}&=\begin{pmatrix}\label{eqn:R4}
-0.221 & 0.108 & 0.108\\
0.108 & -0.221 & 0.108\\
0.108 & 0.108 & -0.221
\end{pmatrix},\\
\bm\sigma^{(4)}&=(-550.8,\ -87.65,\ 212.8,\ 212.8,\ 212.8).\label{eqn:sigma4}
\end{align}
We set the initial distribution as $\bm\lambda^0=(0.256,\,0.144,\,0.128,\,0.206,\,0.267)$.

From (\ref{eqn:R4}), (\ref{eqn:sigma4}), we see that the assumptions (i), (ii), (iv) are not satisfied. Fig. \ref{fig:4} shows the values of $N\mu_i^N$. This example requires a large number of iterations, $N=100000$, to converge. However, the converged value is close to the value of $\sigma^{(4)}_i$ of (\ref{eqn:sigma4}).
\begin{figure}[H]
\begin{center}
\begin{overpic}[width=0.7\columnwidth]{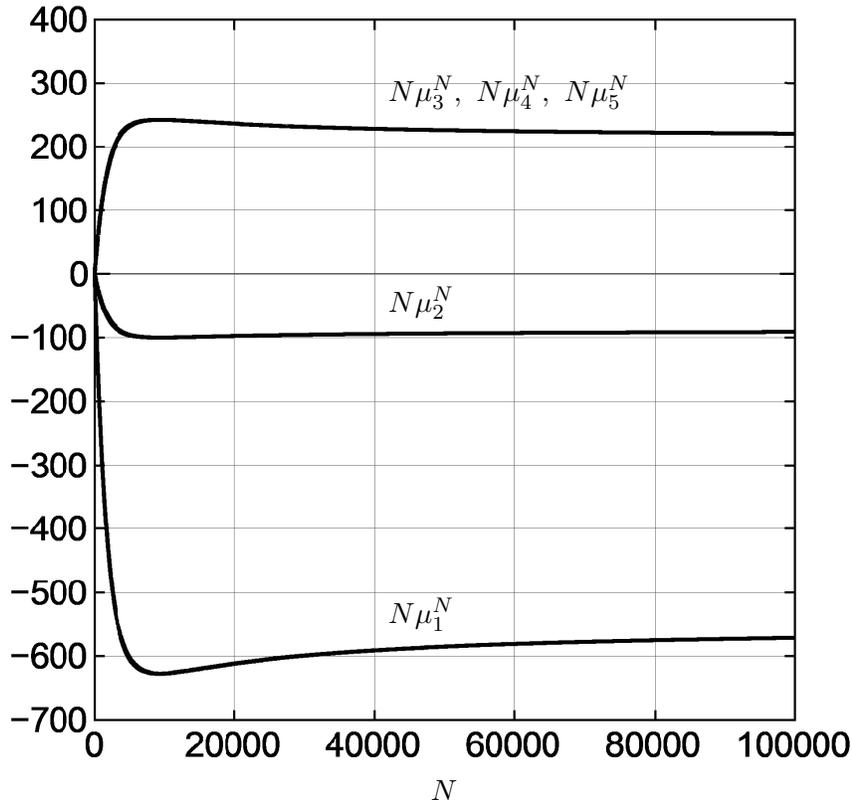}
\put(45,17){$N\mu_1^N$}
\put(45,54){$N\mu_2^N$}
\put(45,79){$N\mu_3^N,\ N\mu_4^N,\ N\mu_5^N$}
\put(50,-4){$N$}
\end{overpic}
\end{center}
\caption{$N\mu_i^N$ of Example 4.}
\label{fig:4}
\end{figure}
\end{example}
\newpage
\section{Conclusion and future issues}
In this study, we considered the convergence speed of the order $O(1/N)$, which is a slow convergence in the Arimoto-Blahut algorithm. We considered the second-order nonlinear recurrence formula consisting of the first- and second-order terms of the Taylor expansion of the defining function of the Arimoto-Blahut algorithm. The convergence of the recurrence formula was proved under the assumptions (i)-(iv). The correctness of the proof was confirmed by several numerical examples. However, the channel matrices satisfying all the assumptions (i)-(iv) were only a very small fraction of the total matrices examined, so further generalization of the proof obtained in this paper is a future issue.
\section{Acknowledgement}This work was supported by JSPS KAKENHI Grant Number JP17K00008.

\end{document}